\documentclass[11pt, a4paper]{article}
\usepackage{jinstpub}
\usepackage[version=4]{mhchem}
\usepackage{placeins}
\usepackage{booktabs}
\usepackage{siunitx}
\usepackage{dsfont}
\usepackage{textcomp}
\usepackage{todonotes}
\usepackage{xspace}
\usepackage{lineno}

\newcommand{\kr}{\ce{^{83\text{m}}Kr}\xspace}
\newcommand{\rb}{\ce{^{83}Rb}\xspace}

\newcommand{\co}{\ce{^{57}Co}\xspace}

\newcommand{\arbeta}{\ce{^{39}Ar}\xspace}
\newcommand{\cevns}{CEvNS\xspace}

\newcommand{\bigO}[1][]{\ensuremath{\mathcal{O}}{#1}}
\newcommand{\cenns}{CENNS-10\xspace}
\newcommand{\csi}[1][ ]{\ce{CsI[Na]}{#1}}

\title{Development of a \kr source for the calibration of the CENNS-10 liquid argon detector}

\newcommand{\itep}{a}
\newcommand{\itepdesc}{\affiliation[\itep]{Institute for Theoretical and Experimental Physics named by A.I. Alikhanov of National Research Centre ``Kurchatov Institute,'' Moscow, 117218, Russian Federation}}

\newcommand{\mephi}{b}
\newcommand{\mephidesc}{\affiliation[\mephi]{National Research Nuclear University MEPhI (Moscow Engineering Physics Institute), Moscow, 115409, Russian Federation}}

\newcommand{\duke}{c}
\newcommand{\dukedesc}{\affiliation[\duke]{Department of Physics, Duke University, Durham, NC 27708, USA}}

\newcommand{\tunl}{d}
\newcommand{\tunldesc}{\affiliation[\tunl]{Triangle Universities Nuclear Laboratory, Durham, NC 27708, USA}}

\newcommand{\utk}{e}
\newcommand{\utkdesc}{\affiliation[\utk]{Department of Physics and Astronomy, University of Tennessee, Knoxville, TN 37996, USA}}

\newcommand{\ornl}{f}
\newcommand{\ornldesc}{\affiliation[\ornl]{Oak Ridge National Laboratory, Oak Ridge, TN 37831, USA}}

\newcommand{\sandia}{g}
\newcommand{\sandiadesc}{\affiliation[\sandia]{Sandia National Laboratories, Livermore, CA 94550, USA}}

\newcommand{\cenpa}{h}
\newcommand{\cenpadesc}{\affiliation[\cenpa]{Center for Experimental Nuclear Physics and Astrophysics \& Department of Physics, University of Washington, Seattle, WA 98195, USA}}

\newcommand{\usd}{i}
\newcommand{\usddesc}{\affiliation[\usd]{Physics Department, University of South Dakota, Vermillion, SD 57069, USA}}

\newcommand{\indiana}{j}
\newcommand{\indianadesc}{\affiliation[\indiana]{Department of Physics, Indiana University, Bloomington, IN, 47405, USA}}

\newcommand{\lanl}{k}
\newcommand{\lanldesc}{\affiliation[\lanl]{Los Alamos National Laboratory, Los Alamos, NM, USA, 87545, USA}}

\newcommand{\laurentian}{l}
\newcommand{\laurentiandesc}{\affiliation[\laurentian]{Department of Physics, Laurentian University, Sudbury, Ontario P3E 2C6, Canada}}

\newcommand{\ncsu}{m}
\newcommand{\ncsudesc}{\affiliation[\ncsu]{Department of Physics, North Carolina State University, Raleigh, NC 27695, USA}}

\newcommand{\virgtech}{n}
\newcommand{\virgtechdesc}{\affiliation[\virgtech]{Center for Neutrino Physics, Virginia Tech, Blacksburg, VA 24061, USA}}

\newcommand{\nccu}{o}
\newcommand{\nccudesc}{\affiliation[\nccu]{Department of Mathematics and Physics, North Carolina Central University, Durham, NC 27707, USA}}

\newcommand{\cmu}{p}
\newcommand{\cmudesc}{\affiliation[\cmu]{Department of Physics, Carnegie Mellon University, Pittsburgh, PA 15213, USA}}

\newcommand{\florida}{q}
\newcommand{\floridadesc}{\affiliation[\florida]{Department of Physics, University of Florida, Gainesville, FL 32611, USA}}

\newcommand{\fermi}{r}
\newcommand{\fermidesc}{\affiliation[\fermi]{Enrico Fermi Institute and Kavli Institute for Cosmological Physics, University of Chicago, Chicago, IL 60637, USA}}

\newcommand{\mipt}{s}
\newcommand{\miptdesc}{\affiliation[\mipt]{Moscow Institute of Physics and Technology, Dolgoprudny, Moscow Region 141700, Russian Federation}}

\newcommand{\tufts}{t}
\newcommand{\tuftsdesc}{\affiliation[\tufts]{Department of Physics and Astronomy, Tufts University, Medford, MA 02155, USA}}

\newcommand{\kaist}{u} 
\newcommand{\kaistdesc}{\affiliation[\kaist]{Department of Physics at Korea Advanced Institute of Science and Technology (KAIST), Daejeon, 34141, Republic of Korea}}

\newcommand{\capp}{v} 
\newcommand{\cappdesc}{\affiliation[\capp]{Center for Axion and Precision Physics Research (CAPP) at Institute for Basic Science (IBS), Daejeon, 34141, Republic of Korea}}

\author[\itep,\mephi]{D.~Akimov,}\itepdesc\mephidesc
\author[\duke,\tunl]{P.~An,}\dukedesc\tunldesc
\author[\duke,\tunl]{C.~Awe,}
\author[\duke,\tunl]{P.S.~Barbeau,}
\author[\utk]{B.~Becker,}\utkdesc
\author[\itep,\mephi]{V.~Belov,}
\author[\utk]{I.~Bernardi,}
\author[\ornl]{M.A.~Blackston,}\ornldesc
\author[\utk]{L.~Blokland,}
\author[\mephi]{A.~Bolozdynya,}
\author[\sandia]{B.~Cabrera-Palmer,}\sandiadesc
\author[\cenpa]{N.~Chen,}\cenpadesc
\author[\usd]{D.~Chernyak,}\usddesc
\author[\duke]{E.~Conley,}
\author[\utk]{J.~Daughhetee,}
\author[\indiana]{M.~del~Valle~Coello,}\indianadesc
\author[\cenpa]{J.A.~Detwiler,}
\author[\cenpa]{M.R.~Durand,}
\author[\utk,\ornl]{Y.~Efremenko,}
\author[\lanl]{S.R.~Elliott,}\lanldesc
\author[\ornl]{L.~Fabris,}
\author[\ornl]{M.~Febbraro,}
\author[\indiana]{W.~Fox,}
\author[\utk,\ornl]{A.~Galindo-Uribarri,}
\author[\laurentian]{A. Gallo Rosso,}\laurentiandesc
\author[\tunl,\ornl,\ncsu]{M.P.~Green,}\ncsudesc
\author[\cenpa]{K.S.~Hansen,}
\author[\ornl]{M.R.~Heath,}
\author[\duke,\tunl]{S.~Hedges,}
\author[\indiana]{M.~Hughes,}
\author[\duke,\tunl]{T.~Johnson,}
\author[\mephi]{A.~Khromov,}
\author[\itep,\mephi]{A.~Konovalov,}
\author[\itep,\mephi]{E.~Kozlova,}
\author[\mephi]{A.~Kumpan,}
\author[\duke,\tunl]{L.~Li,}
\author[\cenpa]{J.T.~Librande,}
\author[\virgtech]{J.M.~Link,}\virgtechdesc
\author[\usd]{J.~Liu,}
\author[\tunl,\ornl]{K.~Mann,}
\author[\tunl,\nccu]{D.M.~Markoff,}\nccudesc
\author[\cenpa]{O.~McGoldrick,}
\author[\ornl]{P.E.~Mueller,}
\author[\ornl]{J.~Newby,}
\author[\cmu]{D.S.~Parno,}\cmudesc
\author[\ornl]{S.~Penttila,}
\author[\duke]{D.~Pershey,}
\author[\ornl]{D.~Radford,}
\author[\cmu]{R.~Rapp,}
\author[\florida]{H.~Ray,}\floridadesc
\author[\duke]{J.~Raybern,}
\author[\itep,\mephi]{O.~Razuvaeva,}
\author[\sandia]{D.~Reyna,}
\author[\fermi]{G.C.~Rich,}\fermidesc
\author[\itep,\mephi]{D.~Rudik,}
\author[\duke,\tunl]{J.~Runge,}
\author[\indiana]{D.J.~Salvat,}
\author[\duke]{K.~Scholberg,}
\author[\mephi]{A.~Shakirov,}
\author[\itep,\mephi,\mipt]{G.~Simakov,}\miptdesc
\author[\duke]{G.~Sinev,}
\author[\indiana]{W.M.~Snow,}
\author[\mephi]{V.~Sosnovtsev,}
\author[\indiana]{B.~Suh,}
\author[\indiana]{R.~Tayloe,}
\author[\virgtech]{K.~Tellez-Giron-Flores,}
\author[\lanl]{R.T.~Thornton,}
\author[\indiana,1]{I.~Tolstukhin,}\note{Now at: Argonne National Laboratory, Argonne, IL 60439, USA}
\author[\indiana]{J.~Vanderwerp,}
\author[\ornl]{R. L.~Varner,}
\author[\ornl]{R. Venkataraman,}
\author[\laurentian]{C.J.~Virtue,}
\author[\indiana]{G.~Visser,}
\author[\cenpa]{C.~Wiseman,}
\author[\tufts]{T.~Wongjirad,}\tuftsdesc
\author[\tufts]{J.~Yang,}
\author[\cmu]{Y.-R.~Yen,}
\author[\kaist,\capp]{J.~Yoo,}\kaistdesc\cappdesc
\author[\ornl]{C.-H.~Yu,}
\author[\indiana,2]{and J.~Zettlemoyer}\note{Now at: Fermi National Accelerator Laboratory, Batavia, IL 60510, USA}
\collaboration{COHERENT collaboration}

\emailAdd{jzettle@fnal.gov}

\abstract{We report on the preparation of and calibration measurements with a \kr source for the \cenns liquid argon detector. \kr atoms generated in the decay of a \rb source were introduced into the detector via injection into the Ar circulation loop. Scintillation light arising from the \SI{9.4}{\keV} and \SI{32.1}{\keV} conversion electrons in the decay of \kr in the detector volume were then observed. This calibration source allows the characterization of the low-energy response of the \cenns detector and is applicable to other low-energy-threshold detectors. The energy resolution of the detector was measured to be 9$\%$ at the total \kr decay energy of 41.5~keV. We performed an analysis to separately calibrate the detector using the two conversion electrons at 9.4~keV and 32.1~keV.}


\begin{document}

\maketitle

%\linenumbers %Turn on line numbers
\section{Introduction}
Coherent elastic neutrino-nucleus scattering (\cevns) was proposed in 1974~\cite{Freedman:1973yd,Kopeliovich:1974mv} following the discovery of the weak neutral current~\cite{bib:Hasert}.
In \cevns, the wavelength of the momentum transfer between the neutrino and the scattering nucleus is larger than the size of the nucleus.
This results in a coherent enhancement to the \cevns cross section but restricts the process to relatively low-energy neutrinos $\bigO(\SI{10}{\MeV})$ for an argon nucleus.
As the nucleus elastically scatters, the only detectable signature is a low-energy $\bigO(\SI{10}{keV})$ nuclear recoil.
The Standard Model provides a precise prediction of the \cevns cross-section and any deviations from the prediction could be indicative of physics beyond the Standard Model. \cevns also provides an opportunity for many interesting physics searches~\cite{Akimov:2020,Zettlemoyer:2020kgh}.

\cevns remained undetected until the COHERENT collaboration first observed the interaction with a 14.6-kg \csi crystal in 2017~\cite{bib:coherentScience}.
A primary goal of COHERENT is to measure the characteristic dependence of the \cevns cross section on the number of neutrons squared ($N^{2}$)~\cite{Akimov:2018ghi}.
To that end, the 24-kg (active), single-phase, liquid argon detector, \cenns, was deployed in the low-background ``Neutrino Alley'' at the Spallation Neutron Source at the US Department of Energy's Oak Ridge National Laboratory (ORNL) in late 2016.
\cenns provides a relatively light target nucleus to begin measuring the $N^2$ dependence of the \cevns cross section. An initial engineering run made measurements of the neutron backgrounds associated with the Spallation Neutron Source beam-on-target and placed a limit on the \cevns cross section~\cite{Akimov:2019rhz}. COHERENT produced the world's-first measurement of \cevns on argon using first data from \cenns~\cite{Akimov:2020,Zettlemoyer:2020kgh}.

\cevns events in liquid argon lie in a range of 5 to 40~keVee, where keVee indicates the reconstructed electron equivalent energy. Thus, \cevns detection with \cenns required a precise characterization of the low-energy detector response. To that end, we prepared and introduced an in situ \kr internal calibration source to allow for a detailed understanding of the detector response near the \cevns energy region of interest. 

\begin{figure}[!htbp]
  \centerline{\includegraphics[width=\textwidth]{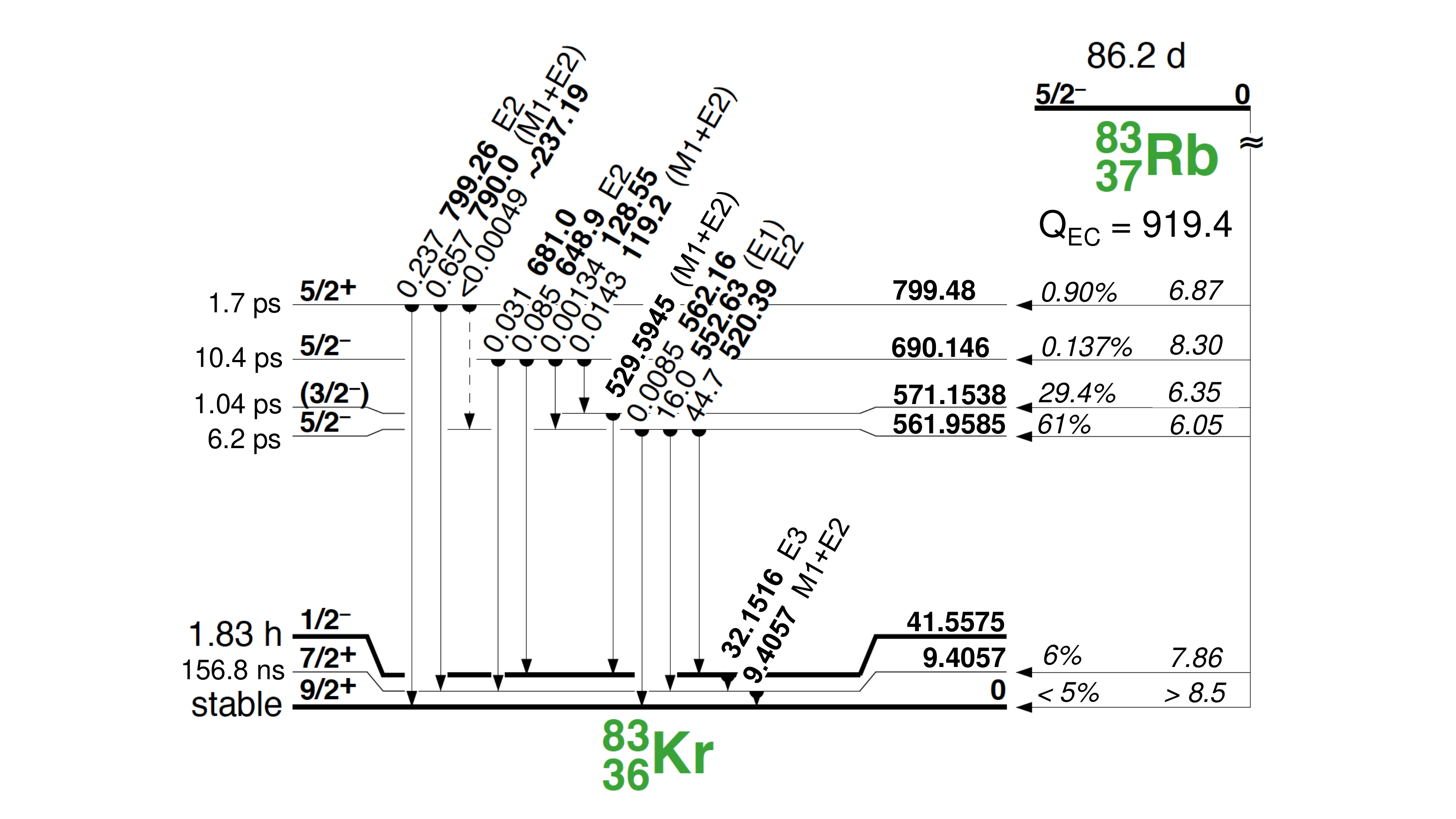}}
  \caption{Energy level diagram of \rb decay to \kr. Figure adapted from scheme taken from Ref.~\cite{bib:tabIsotopes}.}%Zboril et al.~\cite{bib:katrin1}.}
  \label{fig:energy_level}
\end{figure}

\kr is generated through the decay of \rb via electron capture ($t_{1/2} = 86.2$~d). The decay of \rb produces  gamma rays with a range of energies, the most abundant being roughly \SI{500}{\keV}.
These gamma rays can be used to monitor the \kr production rate of the source (see Section~\ref{sec:activity}). Approximately 75\% of \rb decays produce a metastable state of \ce{{}^{83}Kr}, \kr. \kr has an excitation energy of \SI{41.5}{\keV}~\cite{bib:tabIsotopes} (Figure~\ref{fig:energy_level}) and decays to the ground state of \ce{{}^{83}Kr} via internal transition emitting two conversion electrons. The first is emitted with energy \SI{32.1}{\keV} and the second is emitted with a half-life of 157~ns and later with energy 9.4~keV. 

\kr has been used as a low-energy calibration source in noble liquid detectors~\cite{bib:darkside, bib:xenon, bib:lux, bib:kastensXeKr} typically used for direct detection of dark matter, and by experiments designed to directly measure the neutrino mass~\cite{bib:katrin1, Altenmuller:2020}. In the case of direct detection of dark matter, the low-energy nuclear recoil is the same detection signature as \cevns, making \kr a natural choice for the low-energy characterization of \cenns.

We prepared a \kr source and successfully injected it into \cenns. The introduction of this source calibrates the detector electron recoil response to $\sim2\%$ at the \SI{41.5}{\keV} full energy deposition of \kr. Using the light yield from the separate components of the \kr decay, COHERENT calibrated the \cenns detector down to 9.4~keV. This calibration procedure provided a reliable low-energy calibration over a two-year period. In this paper, we describe the preparation of the source, the calibration of \cenns using \kr, the measurement of the source activity using nondestructive radioassay techniques, and the stability of the \kr calibration system.

\section{Preparation of \kr source} \label{sec:app}
A 0.1~M HCl aqueous solution with a total volume of 1.65~mL contained the \rb with an initial activity of 1~mCi. We acquired the \rb solution through the Isotope Business Office at Oak Ridge National Laboratory~\cite{IBO:2020}. We required that the \rb be infused into a matrix that will adsorb the \rb but release the \kr gas atoms to admix with the circulating argon boil-off gas from \cenns. Other liquid-noble-detector applications of \kr sources have successfully used zeolite~\cite{bib:lippincott,bib:kastensXeKr} and activated carbon matrices~\cite{bib:kastens2}. We chose OVC 4x8 activated carbon produced by Calgon Carbon Corporation because of the low radon concentration in activated carbon compared with a zeolite matrix. 

The \kr source preparation procedure broadly followed one used by DarkSide, an experiment to search for dark matter using a liquid argon detector~\cite{AgnesCALIS:2017,Saldanha:2017}. A similar device to the one described in this paper was designed at Yale University~\cite{bib:kastens2} and used in the LUX experiment~\cite{bib:lux} with a replica produced for use in the SCENE experiment~\cite{Cao:2014gns}. Prior to deposition of the \rb solution, the activated carbon was baked in a vacuum oven overnight to remove volatile impurities present in the carbon.

The \rb solution was deposited onto the carbon using a 10~{\micro}L syringe. At the time of the source deposition in April 2018, the $10\pm 5$~{\micro}L deposition volume of the syringe used produced a $3.9\pm1.9$~{\micro}Ci or $140\pm70$~kBq \rb source. A period of 55 days elapsed between the calibration of the \rb solution to 1~mCi and the source deposition. Given the large uncertainty in the efficiency of the \kr atoms to reach the detector volume, we created a high-activity source to ensure that sufficient \kr atoms would reach it. The \rb activity could not be measured via the exact deposition volume at the time of deposition nor the solution transferred to a different container in order to minimize worker exposure to \rb. The available equipment and methods to create the \rb source only allowed for a $\pm50\%$ measurement on the deposition volume; this uncertainty was found from tests using water in place of the \rb solution. Therefore, an in situ measurement of the activity was later performed (Section~\ref{sec:activity}). Note that if the same \rb solution is reused for another deposition the puncture of the bottle with the syringe during the initial deposition causes evaporation of the solution over time.

A Swagelok Vacuum Coupling Radiation (VCR)--based apparatus held the \rb-infused carbon. A 1/2 in. VCR tee held the carbon, which sat inside a structure consisting of stainless steel mesh and fine quartz wool to keep the \rb-infused carbon from shifting inside the container. The introduction of any \rb into the liquid argon volume would generate a long-lived $\gamma$-ray background during normal detector operations. This background is prevented using two 2 {\micro}m sintered metal filters on either end of the VCR tee to keep the carbon particulates from entering the detector volume.

After deposition, the VCR tee directly holding the \rb-infused charcoal was sealed but the system remained open through the sections immediately after the filters shown later in Figure~\ref{fig:krplumbing}. The VCR system was continuously purged with \SI{3}{psig} high-purity (99.999\%) gaseous argon (GAr) continuously flowing through the source exiting through a fume hood. During the purge procedure, the source was heated first for 1.5~h at \SI{60}{\celsius}, then for 1~h at \SI{80}{\celsius}, and finally for 1~h at \SI{100}{\celsius}. This procedure removed impurities and evaporates any \ce{HCl} remaining from the \rb solution.

\section{Determination of source activity} \label{sec:activity}
\begin{figure}[!htbp]
\centering % \begin{center}/\end{center} takes some additional vertical space
\includegraphics[width=.45\textwidth]{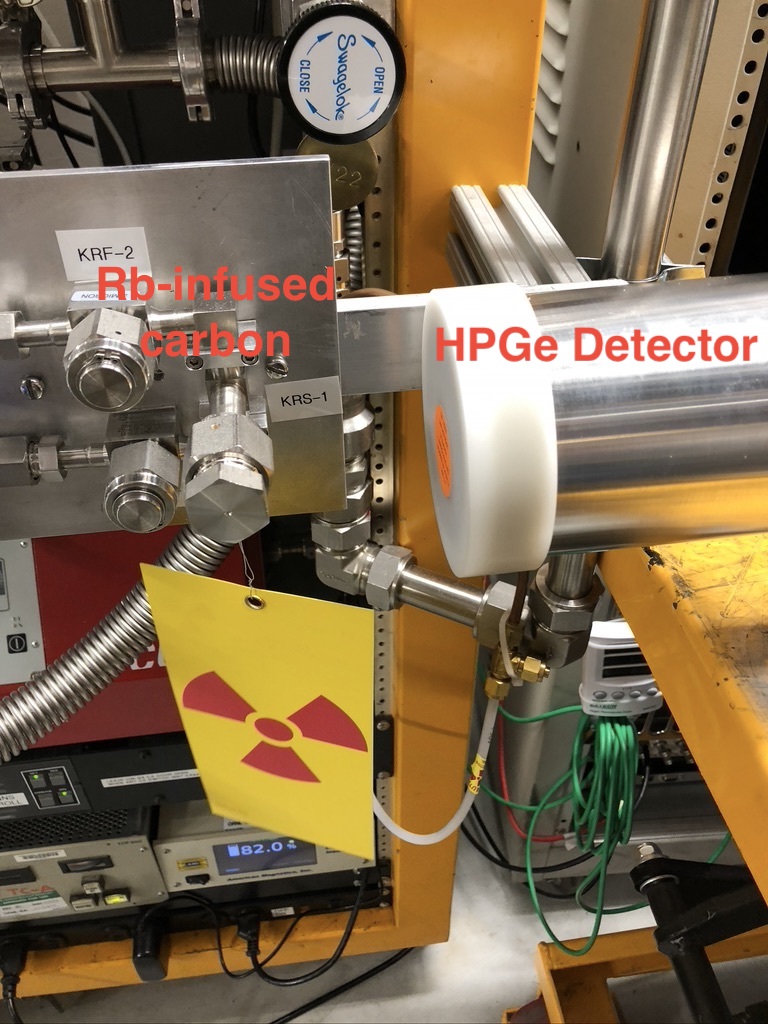}
% "\includegraphics" from the "graphicx" permits to crop (trim+clip)
% and rotate (angle) and image (and much more)
\caption{\label{fig:rbsource} Location of the HPGe detector for the source activity measurement. The detector was placed at a known distance from the \rb source, which was inside the steel VCR container shown in the picture.}
\end{figure}

A high-purity germanium (HPGe) detector, model Canberra BE3820, was used to determine the activity of the \rb source by making a direct gamma spectroscopy measurement generally used in nondestructive assay of nuclear materials. An Inspector 2000 Digital Signal Processor and Multichannel Analyzer was used to acquire the spectra. The Genie2000 spectroscopy software was used as the data acquisition software for the measurement. The measured activity of the \rb source was determined via
\begin{equation} \label{eq:activity}
    A_{\rb} = \frac{N}{\eta'\mu t Y_{\gamma}},
\end{equation}
where $N$ is the total number of counts in the photopeak of interest, $t$ is the measurement time, $\eta'$ is the measured detector efficiency before correcting for the VCR container attenuation, $\mu$ is the VCR container attenuation factor, and $Y_{\gamma}$ is the branching ratio for a given \rb gamma ray. An hour-long measurement ($t = 3600$~s) of the \rb source activity was performed by placing the HPGe detector at a known distance away from the \rb source as shown in Figure~\ref{fig:rbsource}.

\begin{figure}[!htbp]
  \includegraphics[width=.48\textwidth]{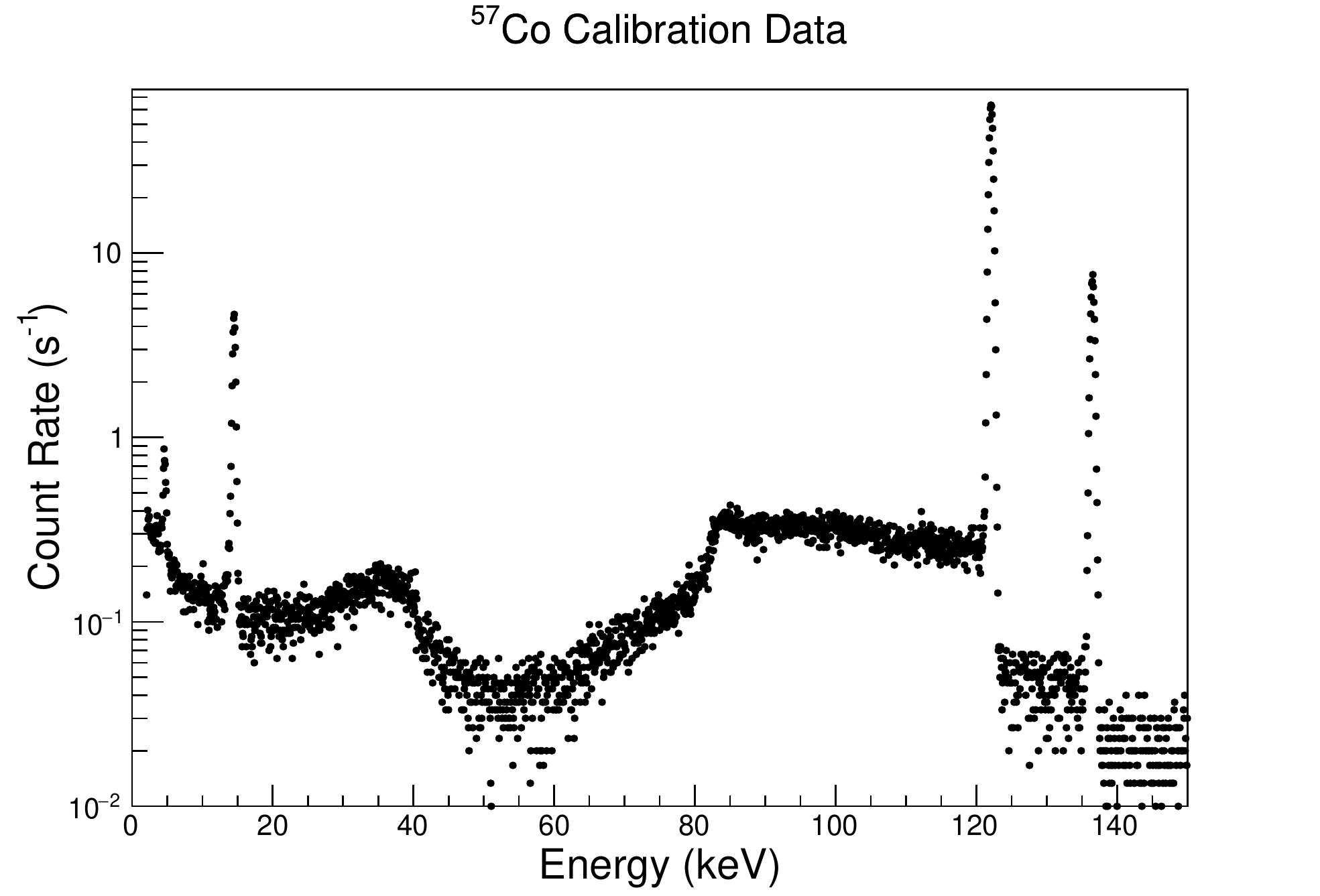}
  \includegraphics[width=.48\textwidth]{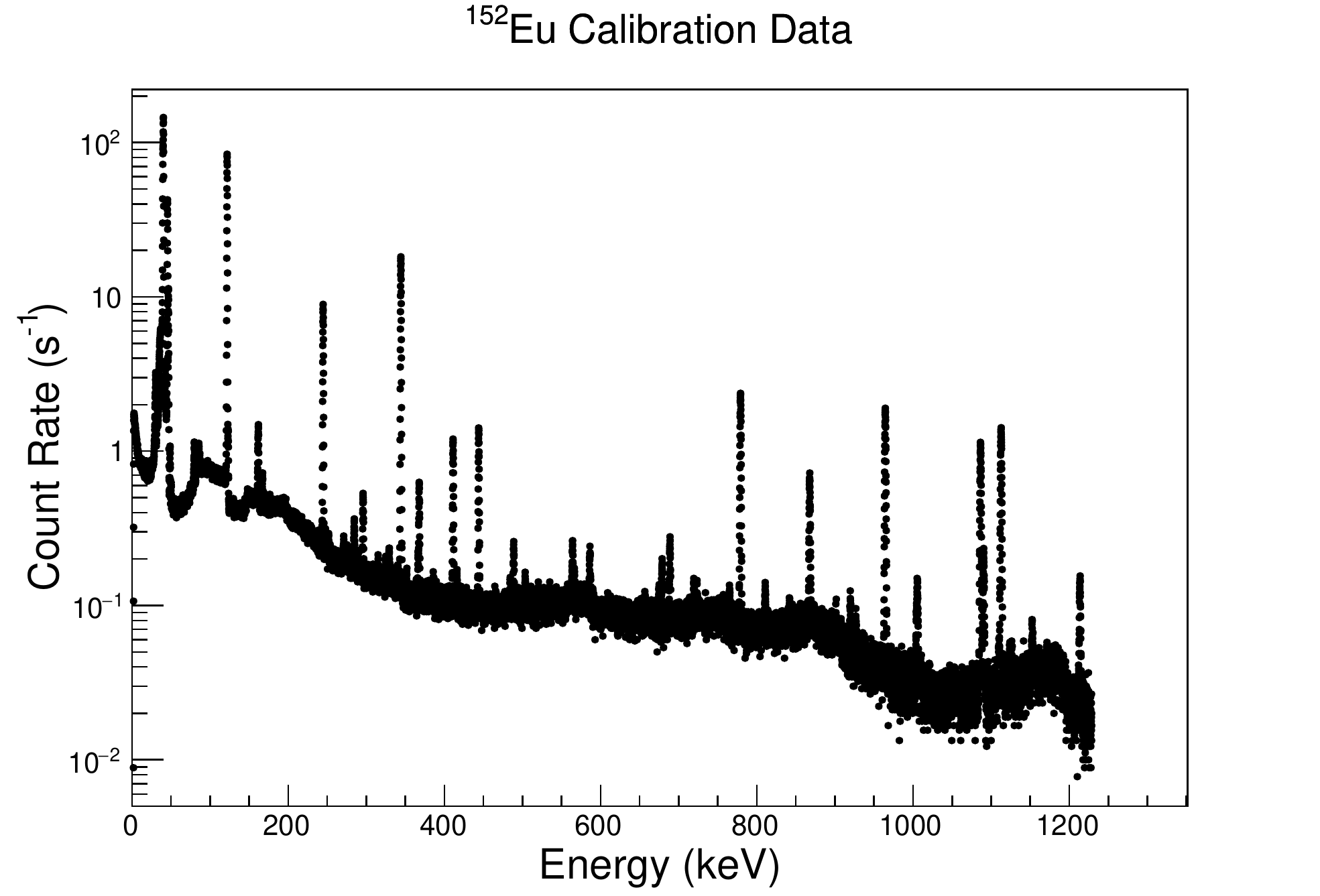}
  \caption{\textbf{Left}: Data from the \co calibration used for the activity measurement. A run length of 300~s was taken with a source with an activity of 27.71 kBq at the time of the measurement. \textbf{Right}: Data from the \ce{{}^{152}Eu} calibration for the activity measurement. A run length of 900~s was taken with a source with an activity of 193~kBq on May 15, 2010}
  \label{fig:ge_calib_spectra}
\end{figure}

To measure the detector efficiency and include the contribution due to container attenuation, we created a replica of the VCR tee that held the carbon. Measurements with \co and \ce{{}^{152}Eu} sources with known activities established the efficiency curve for the detector. Measurements taken with these sources both inside and outside the container were used to characterize the attenuation at the relevant gamma energies due to the VCR source container. The data from the \co and \ce{{}^{152}Eu} sources are shown in Figure~\ref{fig:ge_calib_spectra}.

Uncertainties in the detector efficiency include those on the attenuation from the container, the measured efficiency curve, and the activity of the sources used. In total, these uncertainties produce a $\sim1\%$ error on the final measured activity of the \rb source. The uncertainty in the exact position of the \rb-infused carbon is the major uncertainty on the container attenuation factor. The carbon position uncertainty can be reduced by moving the detector further away from the source although that was not done during this measurement. The absolute activity of the \rb source was computed using Eq.~(\ref{eq:activity}) after the detector efficiency was determined. Results are shown in Table~\ref{tab:rbresults}.

\begin{table}[!htbp]
\centering
\caption{\label{tab:rbresults} Results of the \rb source measurement with the Canberra BE3820 HPGe detector with a measurement time $t=3600$~s.}
\smallskip
\begin{tabular}{l c c c c c}
\toprule
Energy (keV) & $N$ & $\eta'$ & $\mu$ & $Y_{\gamma}$ & $A_{\rb}$(Bq)\\
\midrule
520.4 & \num{30110 \pm 177} & $(4.37\pm0.06)\times10^{-3}$ & $0.83\pm0.01$ & \num{0.450 \pm 0.003} & \num{5145 \pm 79}\\
529.6 & \num{19205 \pm 141} & $(4.29\pm0.06)\times10^{-3}$ & $0.83\pm0.01$ & \num{0.293 \pm 0.002} & \num{5136 \pm 83}\\
552.6 & \num{9902 \pm 103} & $(4.07\pm0.05)\times10^{-3}$ & $0.83\pm0.01$ & \num{0.160 \pm 0.001} & \num{5077 \pm 90}\\
\bottomrule
\end{tabular}
\end{table}
A weighted average of the activity values from Table~\ref{tab:rbresults} gives a final activity of the \rb source of $5.12\pm0.05$~kBq on February 19, 2019, the date of the measurement. The measured value is in agreement with the expected value of $10\pm5$~kBq accounting for the decay of the \rb over the 323~days from the initial deposition to the date of the measurement using the HPGe detector.

\section{Measurements in the CENNS-10 detector}

\subsection{The CENNS-10 detector}
\begin{figure}
  \centerline{\includegraphics[width=0.6\textwidth]{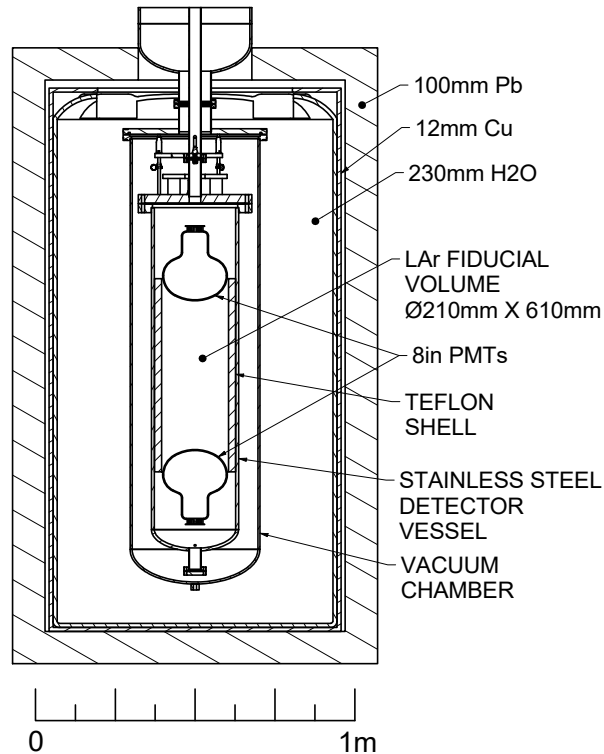}}
  \caption{\cenns detector system. The 24~kg fiducial volume is read out by two Hamamatsu R5912-02MOD PMTs. The shielding structure is designed to minimize backgrounds for a \cevns search~\cite{bib:coherentScience, Akimov:2018ghi}.}
  \label{fig:cenns10_system}
\end{figure}

\cenns is a single-phase scintillation-only liquid argon detector originally designed and built at Fermilab for a \cevns program~\cite{bib:CENNS}.
\cenns was installed at the Spallation Neutron Source as part of COHERENT to measure \cevns in late 2016.
Following the engineering run in the spring of 2017~\cite{Akimov:2019rhz}, \cenns was upgraded to improve the light collection and to increase radiation shielding. Following the upgrade, the detector resumed data-taking for production running in the fall of 2017.
The \cenns detector system as configured for production running is shown in Figure~\ref{fig:cenns10_system}.

In \cenns, the main detector volume is thermally isolated by an insulating vacuum and is suspended from three stainless steel turnbuckles attached to the vacuum vessel lid.
During production running, a cylinder ($r=\SI{20.95}{\cm}$, $h=\SI{60.96}{\cm}$) evaporatively coated with 1,1,4,4-tetraphenyl-1,3-butadiene (TPB) defines the 24~kg active detection region.
TPB is a high-efficiency wavelength shifter~\cite{bib:tpbeff1, bib:tpbeff2} that converts the 128~nm liquid argon scintillation light to the visible ($\lambda_{\mathrm{peak}}\sim400$~nm) where it is more efficiently coupled to the photomultiplier tube (PMT) photocathode response.
TPB-coated PTFE (Teflon) reflecting panels prepared in-house at Oak Ridge National Laboratory define the side walls of the cylinder.
Two frosted-glass Hamamatsu R5912-02MOD cryogenic PMTs evaporatively coated with TPB by Intlvac~\cite{bib:intlvac} were immersed in the liquid argon and provide coverage for detection over the cylinder faces.
The coating thickness was optimized for maximal light detection to an areal density of 0.2~mg/cm$^2$~\cite{bib:mckinsey,lally96}.
The PMT signals were recorded by a 12 bit, 250 MHz digitizer.

The liquid argon scintillation mechanism is well understood.
Energy depositions in liquid argon cause the formation of excited dimer states, which produce 128~nm scintillation light during de-excitation.
These excimers can form into a singlet ($\tau_{1/2}\sim6$~ns) or a triplet ($\tau_{1/2}\sim1.6$~{\micro}s) state, with some indication of an intermediate lifetime~\cite{bib:Hitachi1983}.
The light output of liquid argon strongly depends on the argon purity~\cite{bib:Himi82, bib:WArP10}.
Therefore, the argon in the \cenns system was circulated through a Zr getter to ensure the purity requirements ($\bigO(\SI{1}{ppm})$ \ce{N2} contamination) for scintillation light are met.

This purified gas was then reliquefied in a condenser located above the detector volume, completing the circulation loop.
A Cryomech CP-950 powered a PT-90 cold head to liquefy the GAr, which was precooled after passing through a heat exchanger with the detector boil-off gas.

The \cenns shielding structure was designed to minimize the effects of gamma ray and neutron backgrounds for a \cevns search and is shown in Figure~\ref{fig:cenns10_system}.
The outermost shielding element consists of \SI{10.16}{\cm} of chevroned lead bricks.
Inside the lead is \SI{1.27}{\cm} of copper to minimize background from bremsstrahlung radiation due to \ce{{}^{210}Pb} decays.
Finally, the innermost shielding element consists of \SI{23}{\cm} of water to minimize backgrounds from external neutron sources.

\subsection{Modes of operation} 
\begin{figure}[!htbp]
  \centerline{\includegraphics[width=0.99\textwidth]{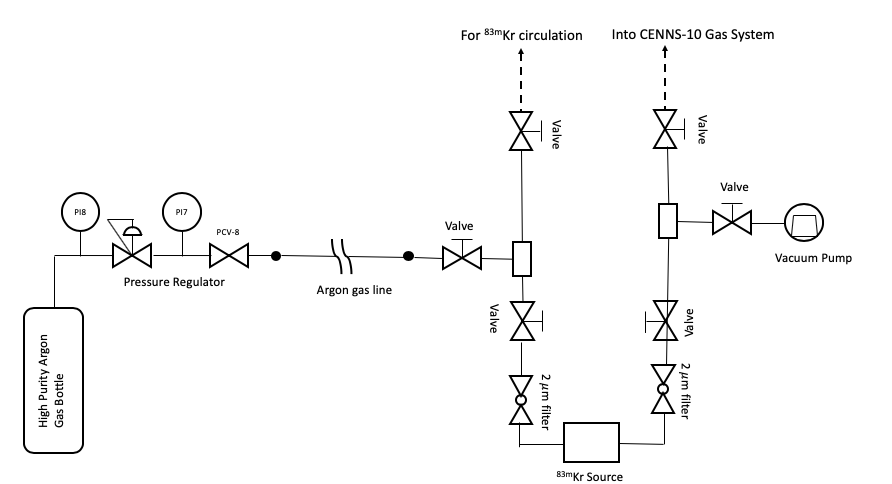}}
  \caption{\cenns \kr source plumbing diagram. The source attaches directly to the \cenns gas-handling system. If necessary, the source can be bypassed to flow GAr directly into the detector. The \kr source is contained in a 1/2 in. VCR cap and tee holding the \rb-infused carbon.} 
  \label{fig:krplumbing}
\end{figure}

The \kr calibration system was designed to be able to operate in either ``injection'' or ``circulation'' mode depending on the position of valves located throughout the system. Figure \ref{fig:krplumbing} shows the plumbing diagram, including valves associated with the introduction of the \kr source. All valves shown were open to operate in injection mode with the exception of those leading to the vacuum pump and the circulation inlet. In injection mode operation, the \kr atoms were introduced into the system with a one-time injection. After evacuating the \kr source with a turbomolecular pump, \kr gas was allowed to accumulate. The initial evacuation of the system was optional but was used as a method to control the resulting rate of \kr decays that depend on the length of the accumulation time after evacuating. The \kr was then introduced into the system by flowing \SI{99.999}{\percent} argon gas from a ultrahigh-purity bottle through the source and into the \cenns circulation loop. 

In circulation mode, the \kr system was directly introduced into the \cenns GAr circulation loop to allow a constant flow of argon to pass through the \kr system and the charcoal. This allowed for the rate of \kr decays seen in the detector to reach a steady-state value dependent on the activity of the \rb source. Although the design allows for operation in circulation mode, this bypasses a flow meter that finely controls the flow rate of the recirculating GAr and introduces instability of \cenns detector operations. Therefore, this mode was not used for calibration. Future improvements to the system for circulation mode operation will stabilize the \kr concentration in \cenns and better inform the \kr transport efficiency into the active volume. The improvements will require a redesign of the gas handling system and are being considered for a new detector.

\begin{figure}[!htbp]
\includegraphics[width=.48\textwidth]{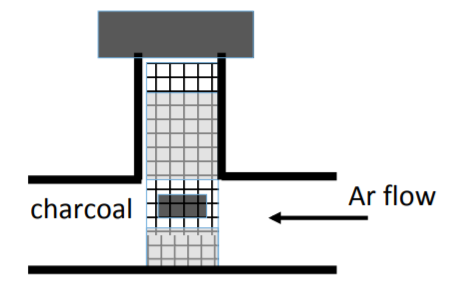}
\includegraphics[width=.48\textwidth]{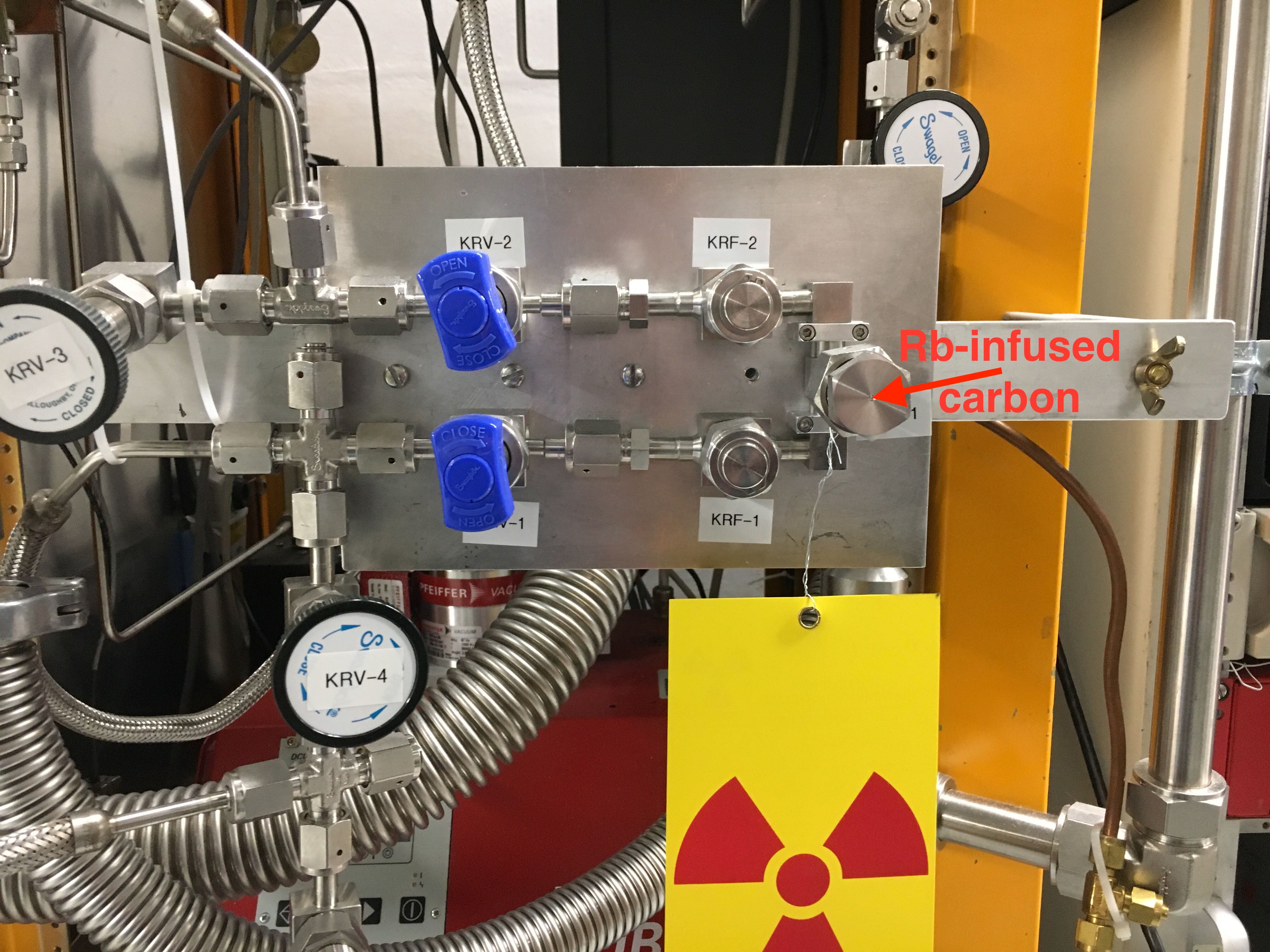}
    \caption{\textbf{Left}: Diagram of carbon location within the VCR tee. A structure of thin steel mesh and fine quartz wool holds the carbon in place within the tee. \textbf{Right}: \kr source attached to the \cenns gas-handling system in operation mode focused on the VCR tee holding the carbon and surrounding parts. The VCR tee is the section in the center on the right side of the source.}
    \label{fig:carbon}
\end{figure}

The design of the \kr source apparatus allowed for GAr to flow through or bypass the \rb-infused carbon vessel on its path into the \cenns detector. The apparatus was also connected to a turbomolecular pump and could be evacuated to check for vacuum leaks. The normal recirculation path of the boil-off GAr from the \cenns volume carried the \kr atoms through the \cenns gas handling system, the liquefier, and into the active volume. A diagram of the location of the source inside the VCR tee attached to the \cenns gas-handling system is shown in Figure~\ref{fig:carbon}.

\subsection{Tests prior to injection}
Before introducing the \kr-activated carbon, we performed tests of the apparatus in injection and circulation modes by attaching the VCR tee to the \cenns gas-handling system. The two tests included the VCR tee system both with and without clean carbon inside. These tests using a \co calibration source showed that no degradation of the detector light collection efficiency occurred and no additional impurities were introduced from adding GAr in injection mode. The injection mode allowed the detector to remain operational where the recirculation rate is finely controlled and the recirculating gas passes through the Zr getter. We first observed the gas-system instability associated with circulation mode operation during these tests. 

We performed several runs in April 2018 to August 2019 using the \kr calibration system described in Section~\ref{sec:app}, usually during normal Spallation Neutron Source shutdown periods. The continued runs over a lengthy period demonstrate the capability to repeatedly calibrate the \cenns detector with a single \kr source. 

For the initial runs, we evacuated the \kr source using the turbomolecular pump to remove the built-up \kr gas during the period between runs. After the evacuation, we allowed the \kr gas to accumulate over a period of approximately 2 h. To achieve the maximal rate during an injection due to the reduced source activity in later runs, the source region was not evacuated before the injection of \kr.  

\subsection{Calibration of the \cenns detector with \kr}
For the \kr calibration, the \cenns data acquisition was triggered by the coincidence of two PMT signals above a voltage threshold within a 24~ns window. The set threshold was above the single photoelectron (SPE) pulse height in the PMTs to prevent random coincidences. A pulsed LED calibration measured the approximate height of the SPE pulses. 

\begin{figure}[!htbp]
\centering % \begin{center}/\end{center} takes some additional vertical space
\includegraphics[width=.48\textwidth]{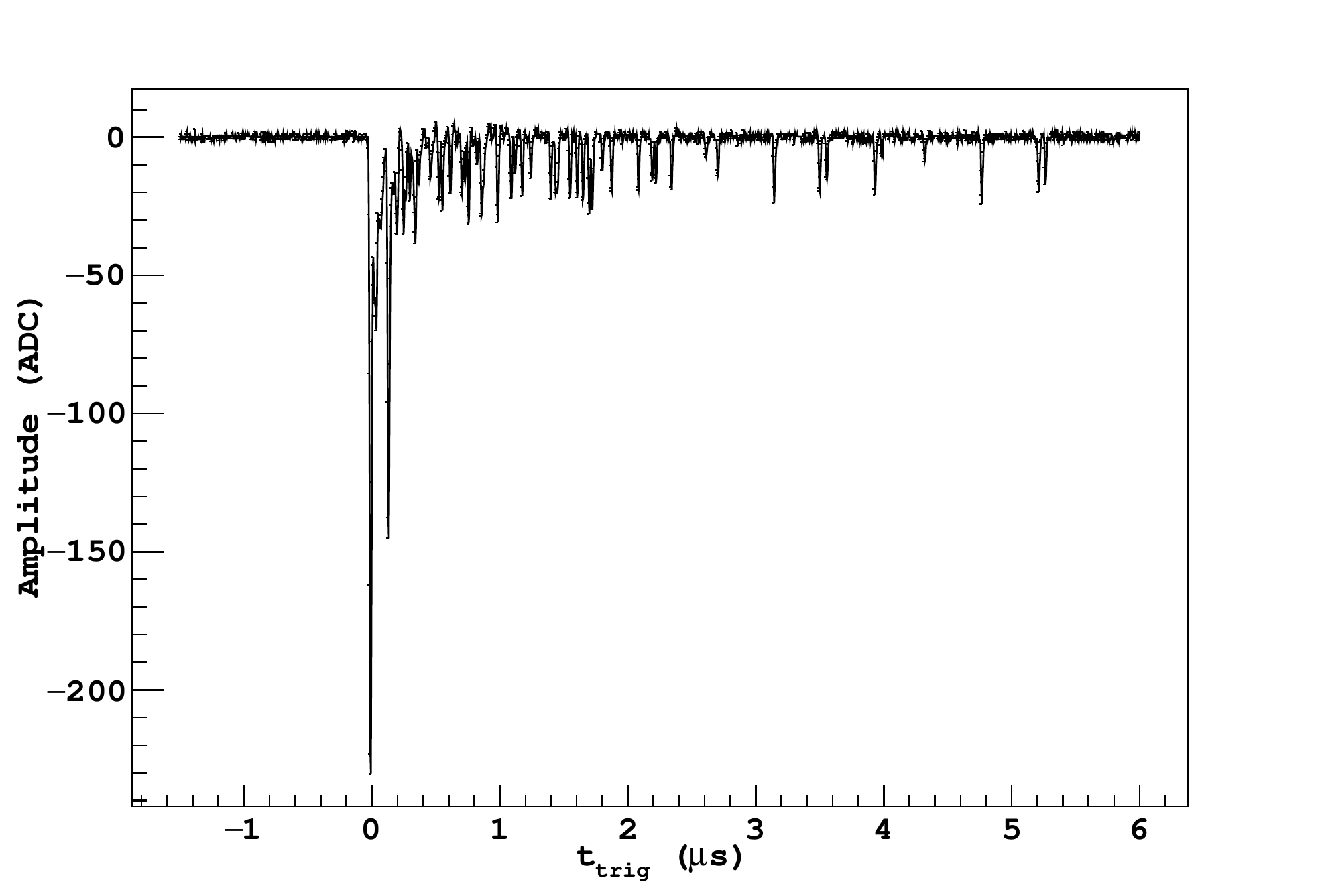}
\includegraphics[width=.48\textwidth]{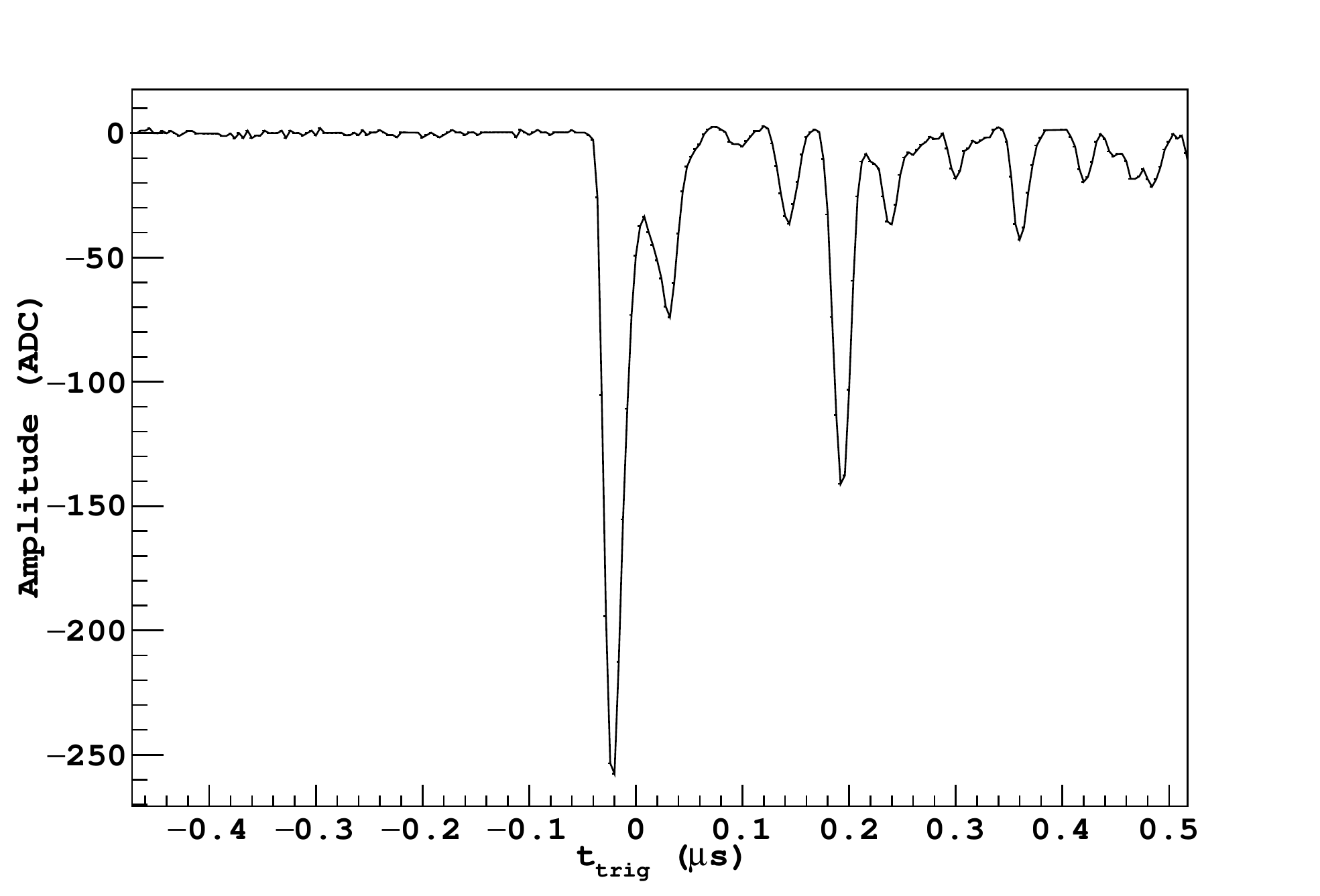}
% "\includegraphics" from the "graphicx" permits to crop (trim+clip)
% and rotate (angle) and image (and much more)
\caption{\label{fig:krevent} \textbf{Left}: Example of \kr event waveform from one of the \cenns PMTs. The large pulse at $\sim0$~{\micro}s is the prompt light from the \SI{32.1}{\keV} component and the second clear peak at $\sim0.2$~{\micro}s is the prompt light from the \SI{9.4}{\keV} component. The triplet light from the two components is indistinguishable in the data acquisition window. \textbf{Right}: The same event zoomed into early times to clearly show the second 9.4~keV component pulse separated from the initial pulse.}
\end{figure}

The digitized waveforms were recorded for a range of $-1.5$~{\micro}s $<t_{\mathrm{trig}}<6$~{\micro}s where $t_{\mathrm{trig}}$ is the time of the waveform trigger, which is the initial pulse. The waveforms were corrected for known capacitive coupling effects and pulses were identified from the corrected waveforms using a voltage-level threshold. PMT waveforms are further required to have a stable baseline and no ADC saturation. The total number of photoelectrons (PEs), $I$, was extracted from the corrected waveform in a 6~{\micro}s window after the initial pulse near $t_{\mathrm{trig}}=0$. For \kr data, this choice of integration window combines the two depositions into a calibration measurement at 41.5~keV. The long lifetime of the triplet state of liquid argon scintillation drove the choice of integration window. Candidate events were then formed by requiring pulses with $\ge2$~PE in both PMTs occurring within $20$~ns of each other. An example \kr event waveform is shown in Figure~\ref{fig:krevent}.

The \kr source produced a clear peak in the energy spectrum above the steady-state backgrounds seen in the detector, which corresponds to the full \SI{41.5}{\keV} energy deposition. Background spectra were taken during these runs before the introduction of \kr to allow for a background subtraction of the \kr data. After a background subtraction was performed, the resulting \kr peak at \SI{41.5}{\keV} was fit to a Gaussian distribution and the energy resolution was measured to be 9.3\% at \SI{41.5}{\keV}, or approximately $\frac{1.3}{\sqrt{N}}$ where $N$ is the number of measured PEs. We measured the energy mean uncertainty to $\sim2\%$, combining errors on the integrated SPE, the spread of calibration results in the PE space, and the Gaussian fit model for the \kr calibration source. 

\begin{figure}[!htbp]
\centering % \begin{center}/\end{center} takes some additional vertical space
\includegraphics[width=.48\textwidth]{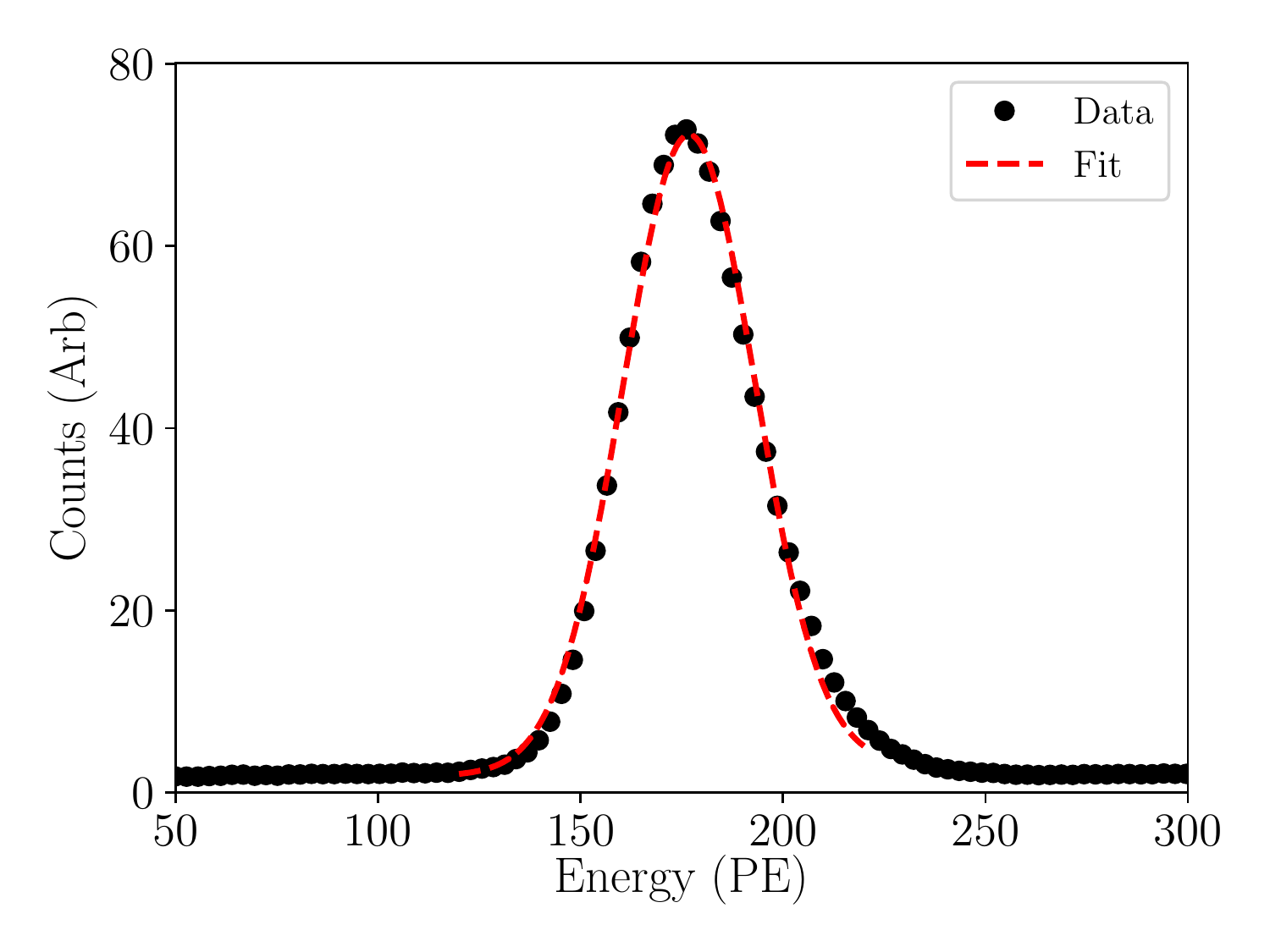}
\includegraphics[width=.48\textwidth]{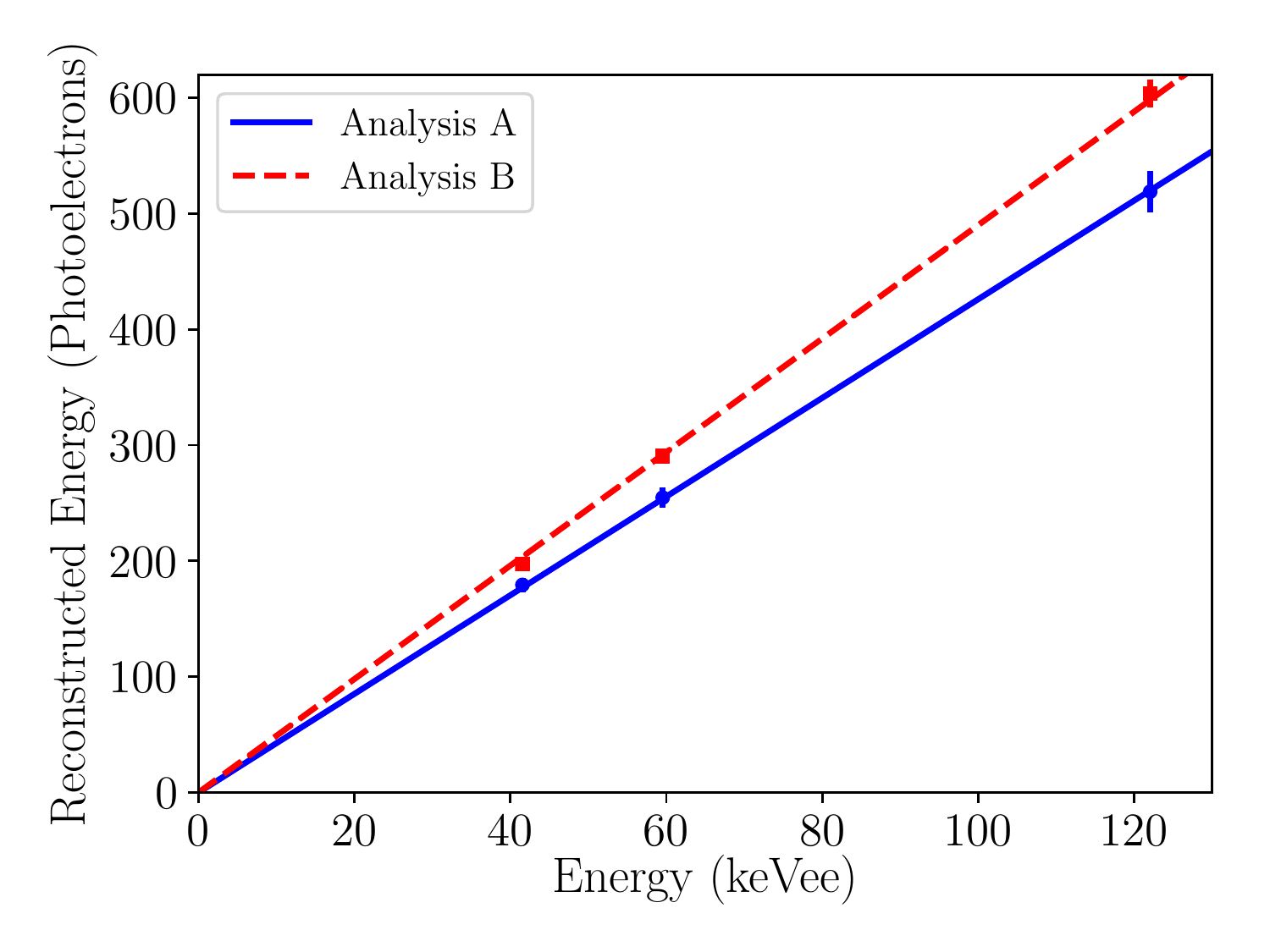}
% and rotate (angle) and image (and much more)
\caption{\textbf{Left}: Reconstructed \kr event energy spectrum from a single calibration run. There is a clear peak at $\sim$177~PE corresponding to the full \SI{41.5}{\keV} energy deposition due to the \kr decays. The data are fit to a Gaussian distribution (dashed line) with $\sigma/E=9.3\%$. The excess in the tails of the data compared with the fit is due to events occurring near the PMTs. \textbf{Right}: Results of the calibration of the \cenns detector with \co (\SI{122}{\keV}), \ce{{}^{241}Am} (\SI{59.5}{\keV}), and \kr (\SI{41.5}{\keV}) sources. Using these sources, the linearity of the detector response is established in the range of \SIrange{41.5}{122}{\keV} with a light yield of $Y=4.3\pm0.1$~~PE/keV from Analysis A in blue and $Y=4.9\pm0.1$~PE/keV from Analysis B in red.}
\label{fig:krspectrum}
\end{figure}

Figure~\ref{fig:krspectrum} shows an example reconstructed energy spectrum in units of PEs from one of the \kr source runs. The exact location of the \kr event peak in this plot depends on the definition of the SPE used within an analysis. Two parallel analyses of the \cenns data, labeled ``Analysis A'' and ``Analysis B'' and described further in Ref.~\cite{Akimov:2020}, choose different definitions of the integrated SPE resulting in different values for the light yield, $Y$, of the detector. There is one key difference in the determination of the value of the integrated SPE. Analysis A adopts a model where the integrated SPE is extracted from the fully amplified SPE component with the fit function described further in Ref.~\cite{DEAPSPE:2017} fitting to LED calibration data. Analysis B adopted a single Gaussian model to describe the integrated SPE fitting the late triplet light from source calibration data within a small window. The two analyses independently measured a very similar mean energy uncertainty and therefore, this choice only minimally affects the full reconstruction from digitized waveform to reconstructed energy.

External \ce{{}^{241}Am} and \co sources along with this \kr source established the linearity of the \cenns detector response in the range of \SIrange{41.5}{122}{\keV}. As seen in the right panel of Figure~\ref{fig:krspectrum}, a linear detector response and a light yield of $Y=4.3\pm0.1$~PE/keV was observed from Analysis A and $Y=4.9\pm0.1$~PE/keV from Analysis B. The difference in the measured values is almost exclusively due to the different definition of the SPE. The important result is that both analyses measure a linear detector response using the same set of source measurements. The value of the location in PEs of the \kr peak used in Figure~\ref{fig:krspectrum} is an average of the various runs performed using the \kr source.

\begin{figure}[!htbp]
  \includegraphics[width=.48\textwidth]{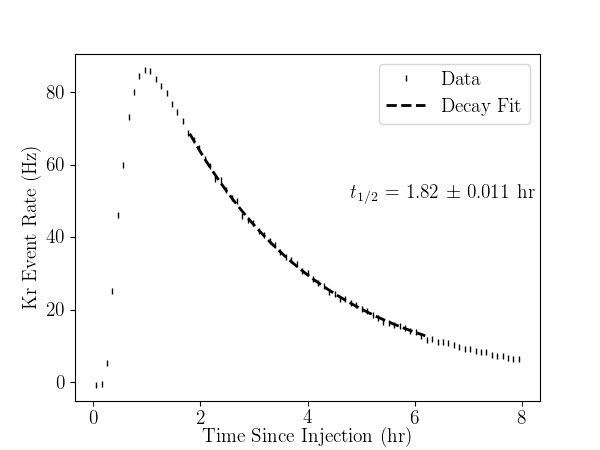}
  \includegraphics[width=.48\textwidth]{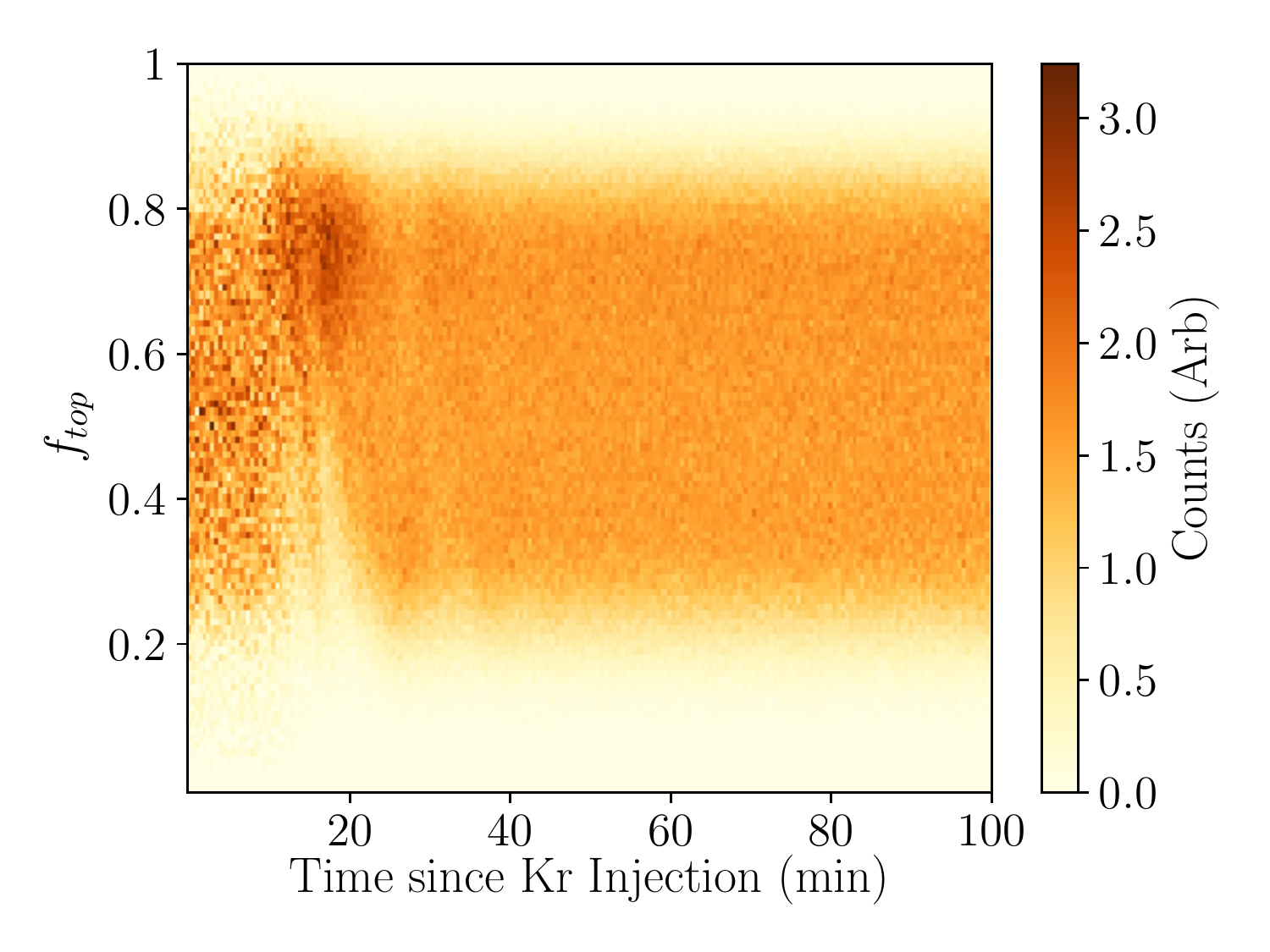}
  \caption{\textbf{Left}: Krypton event rate vs. time after a \kr injection through the argon circulation system. The event rate decays after a maximum value soon after the injection occurs. The best-fit half-life of the decay is $1.82\pm0.01$~h. This value is in good agreement with the literature value of 1.83~h. \textbf{Right}: Distribution of $f_{top}$ over time during the \kr calibration. Events are concentrated near the top PMT at early times. The initial $\sim$10 min represent data taken before the \kr was introduced into the system. After roughly 30~min, the \kr becomes evenly distributed within the detector volume.}
  \label{fig:krcircrate}
\end{figure}

Figure~\ref{fig:krcircrate} shows the \kr event rate as a function of time during an injection mode run. Data were taken with the \kr source for a period of time after the injection occurred until the event rates returned to steady-state background levels. Observing the time-evolution of the \kr event rate during the injection mode run, we found a best-fit half-life of $1.82~\pm0.01$~h, in agreement with the reported half-life of 1.83~h in~\cite{bib:tabIsotopes}. 

The right sub-figure in Figure~\ref{fig:krcircrate} shows the distribution of \kr events throughout the detector volume by examining the fraction of total light seen by the top PMT, $f_{\mathrm{top}}$, defined as
\begin{equation} \label{eq:asymm}
    f_{\mathrm{top}} = \frac{I_{\mathrm{top}}}{I},
\end{equation}
where $I_{\mathrm{top}}$ is the integrated signal for the top PMT and $I$ is the total integrated signal from the top and bottom PMTs. Examination of the time dependence of this parameter shows a clear concentration of higher $f_{\mathrm{top}}$ values at the beginning of the run before a uniform distribution in the \cenns detector volume occurs. This indicates that the \kr source events are more concentrated toward the top PMT when the \kr gas is first introduced into the system but evenly distributed throughout the detector volume at later times.

During a run in which the measured rate was allowed to reach the maximal value, the observed \kr rate of 2.6~kBq was 68\% of the expected rate of 3.8~kBq at the time of the run. The expected rate was determined from the measurement in Section~\ref{sec:activity}, accounting for the fiducial volume of the detector and the branching ratio of \rb decays that produce \kr (Figure \ref{fig:energy_level}). Although the exact reason for the discrepancy in the measured rate compared with the expected rate is unknown, one possibility is the unaccounted freeze-out of \kr atoms on detector surfaces. This fraction of detected events still ensures a significant number of \kr decays in the \cenns volume over the course of a single run.

\subsection{Component isolation analysis} \label{sec:comp}
As the total energy deposition for a given \kr decay is composed of two components at energies of 9.4~keV and \SI{32.1}{\keV}, one can obtain separate calibrations at each energy. In each decay of \kr, a prompt electron from the 32.1~keV transition is followed by the electron from the 9.4~keV transition with a half-life of 157~ns. In practice, the average light yield of these depositions is difficult to determine through standard integral analysis methods because of the large degree of overlap in the scintillation light arising from the triplet state. Two separate analyses were performed to extract the light yield from each component of the \kr decay using the methods of both Analysis A and Analysis B from Ref.~\cite{Akimov:2020}.

\subsubsection{Analysis A} \label{sec:compA}
\begin{figure}[!htbp]
\centering % \begin{center}/\end{center} takes some additional vertical space
\includegraphics[width=.6\textwidth]{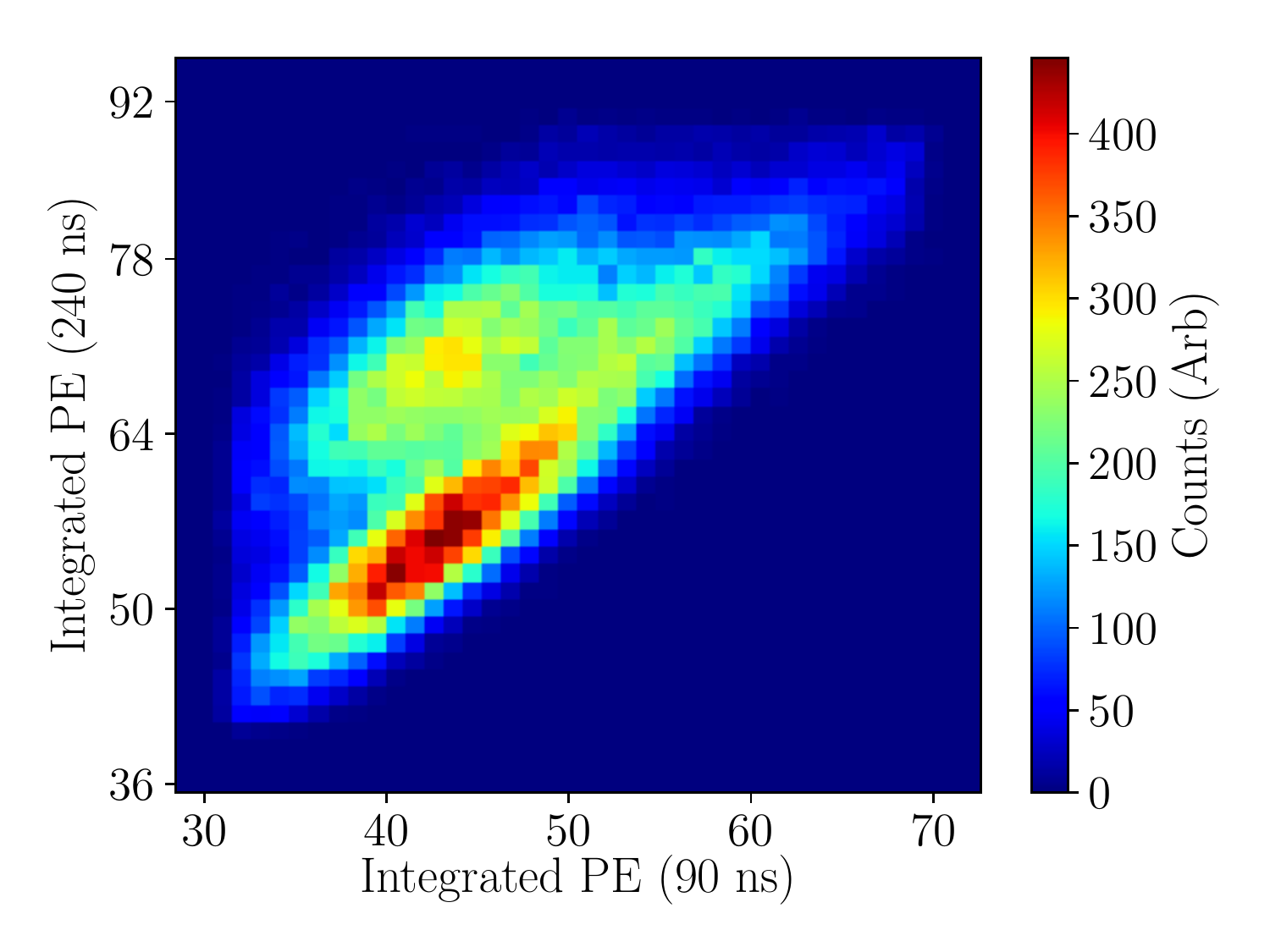}
\caption{\label{fig:IntegralComparison} Integral of \kr event waveforms over 90~ns vs. 240~ns with respect to event onset. The secondary population of events in the upper left represents decays where the \SI{9.4}{\keV} emission occurs after 90~ns but before 240~ns.}
\end{figure}

The first method uses the waveform analysis techniques of Analysis A described further in Ref.~\cite{Akimov:2020}, which directly integrates the event waveform. For any given \kr calibration data set, a population of events with a delayed ($>90$~ns) \SI{9.4}{\keV} energy deposition exists. This population is readily apparent in the data when comparing the integrals of the recorded waveforms at two distinct time intervals (Figure~\ref{fig:IntegralComparison}).

\begin{figure}[!htbp]
  \includegraphics[width=.48\textwidth]{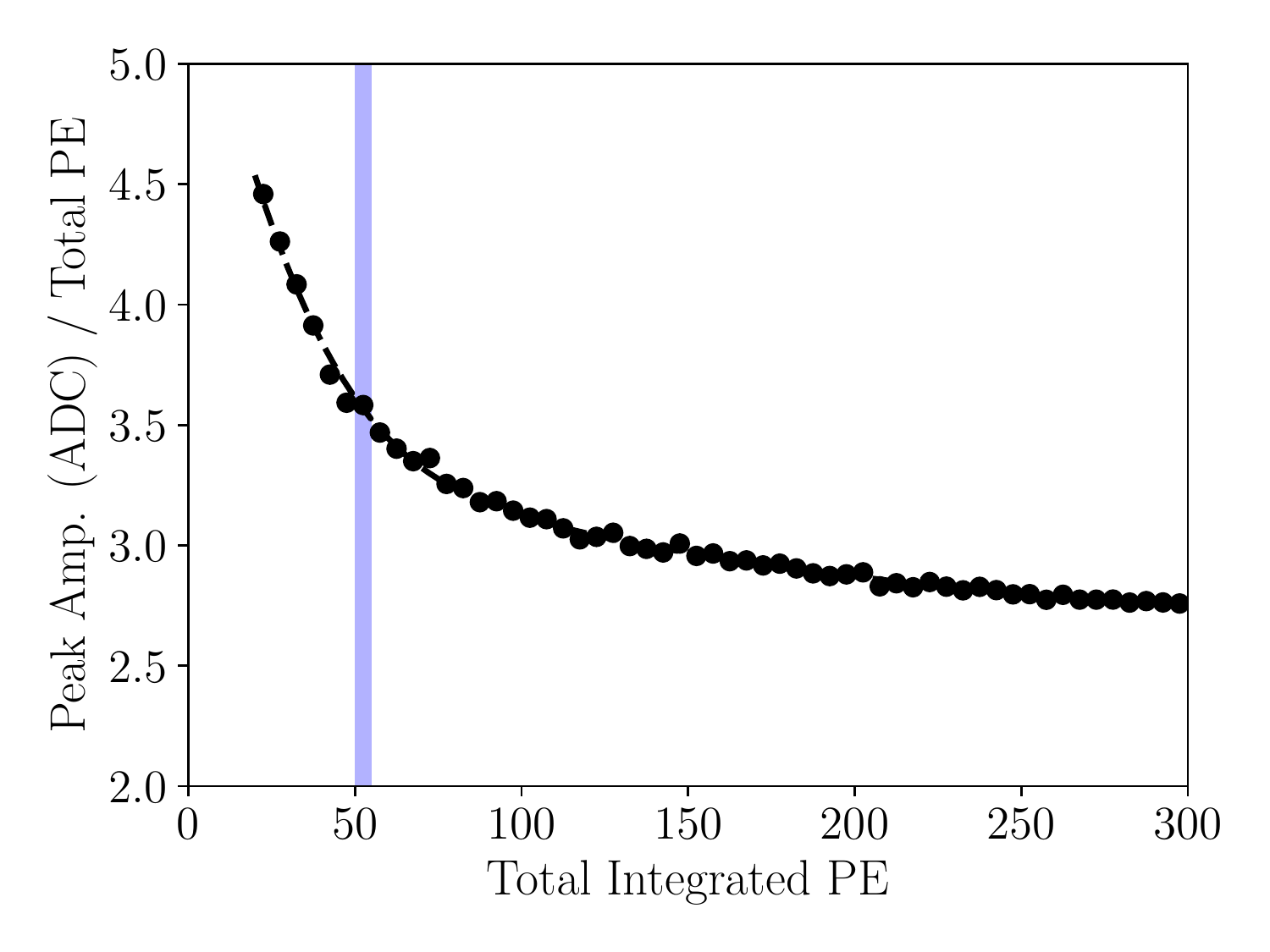}
  \includegraphics[width=.48\textwidth]{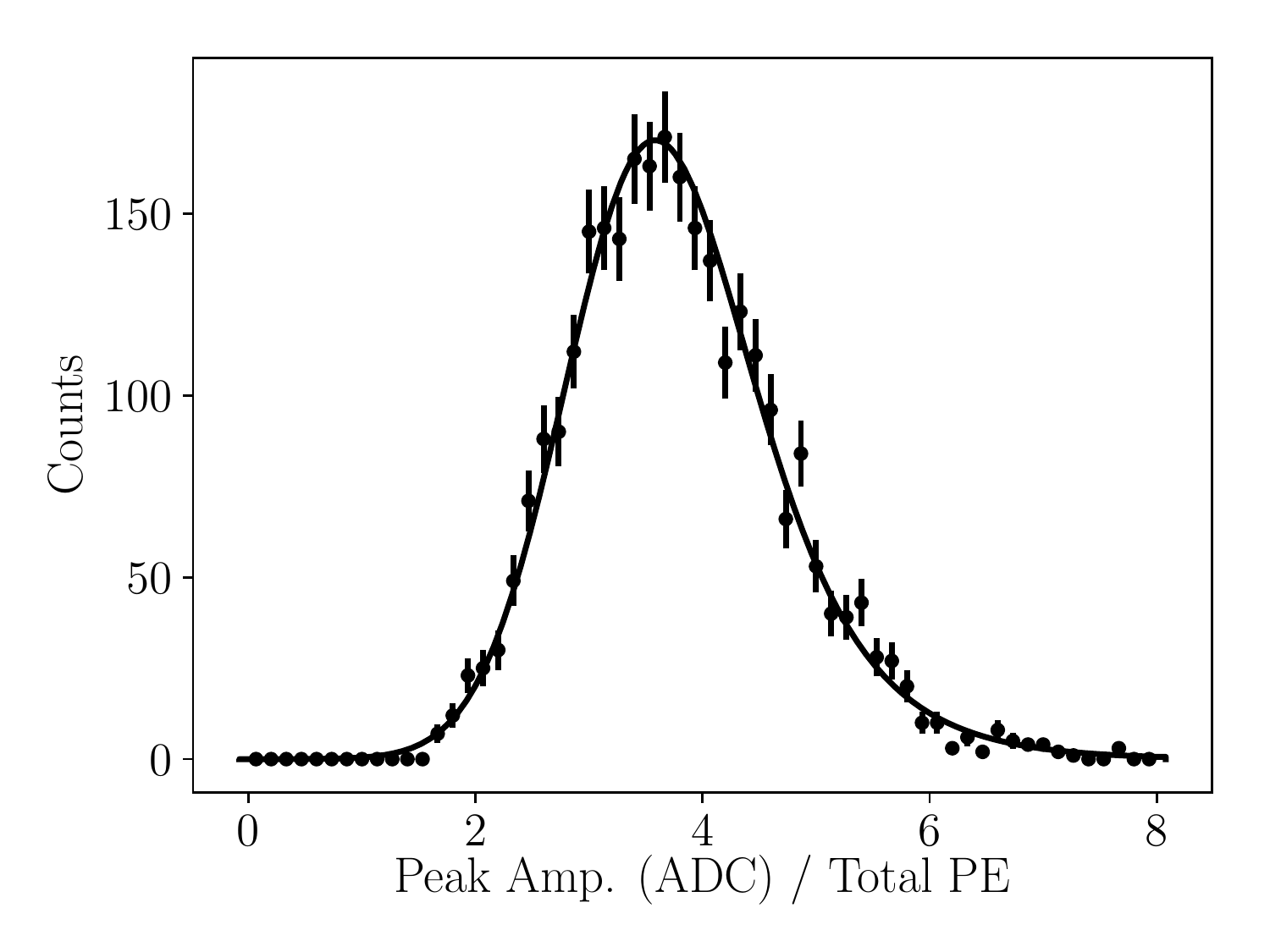}
  \caption{\textbf{Left}: $S_{A_{p}}$ as a function of total integrated PE for electron recoil events. The relationship is derived from background \arbeta events. \textbf{Right}: Distribution of peak-to-total ratio for events in highlighted energy slice.}
  \label{fig:PeakToTotal}
\end{figure}

If the energy depositions are sufficiently separated in time, then it is possible to perform an independent measurement of the light yield using distributions of the respective singlet peak heights. Using a large sample of background \arbeta beta-decay events, one can construct an energy-dependent relationship for the ratio of the singlet peak amplitude in ADC units to the total PE collected for electron recoil events in the \cenns detector (Figure~\ref{fig:PeakToTotal}). The light yield $Y$ for a given energy deposit $E_{\mathrm{dep}}$ is related to the peak amplitude $A_{p}$ via
\begin{equation} \label{eq:lyspline}
E_{\mathrm{dep}} \cdot Y = \frac{A_{p}}{S_{A_{p}}},
%\hspace{-3.0pt}(Y \cdot E_{dep})}
\end{equation}
where $S_{A_p}$ is a function that varies with the value of $Y\cdot E_{\mathrm{dep}}$ that returns the nominal value of the peak-to-integral ratio for the total expected PE of that event. Given the measured $S_{A_{p}}$ for each $A_{p}$, we can solve Eq.~\ref{eq:lyspline} implicitly for $Y$.

To measure the mean peak amplitude for \SI{32.1}{\keV} depositions, we selected events with the \SI{9.4}{\keV} component delayed by at least $152$~ns (preventing any contribution to the initial event peak). A Gaussian fit was then applied to the distribution of the largest amplitude sample in the added waveform of the top and bottom PMTs. The expected total number of PEs for the deposition can then be solved for using the aforementioned peak-to-integral ratio function Eq.~(\ref{eq:lyspline}).

\begin{figure}[!htbp]
\centering
\includegraphics[width=.8\textwidth]{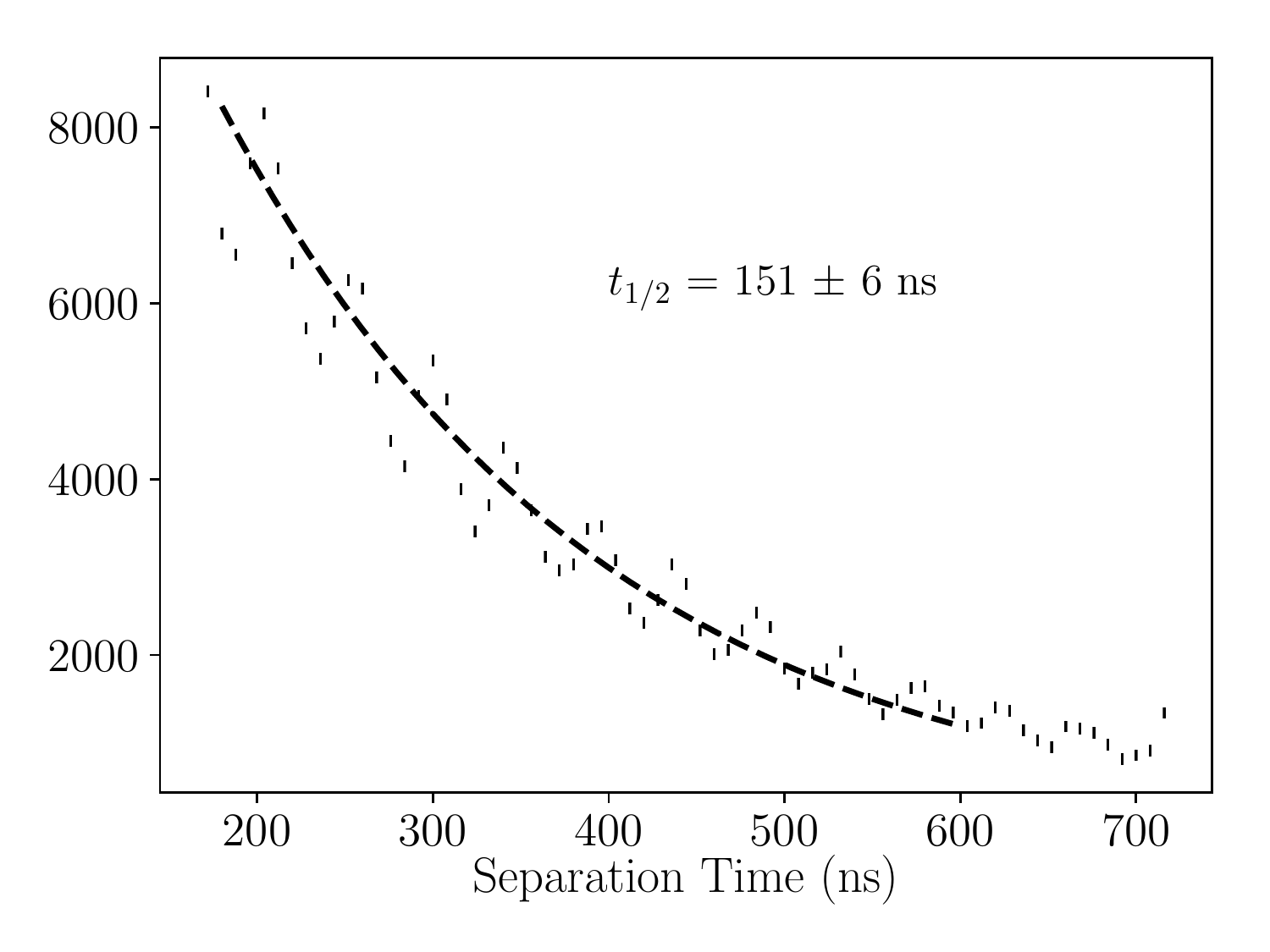}
\caption{Distribution of time between the initial peak representing the singlet scintillation from the 32.1~keV component and the delayed peak, representing the singlet scintillation from the 9.4~keV component. The measured half-life of the decay over the time range examined is $151\pm6$~ns, in agreement with the expected half-life for the 9.4~keV component of 157~ns.}
    \label{fig:halflifetwo}
\end{figure}

\begin{figure}[!htbp]
\includegraphics[width=.48\textwidth]{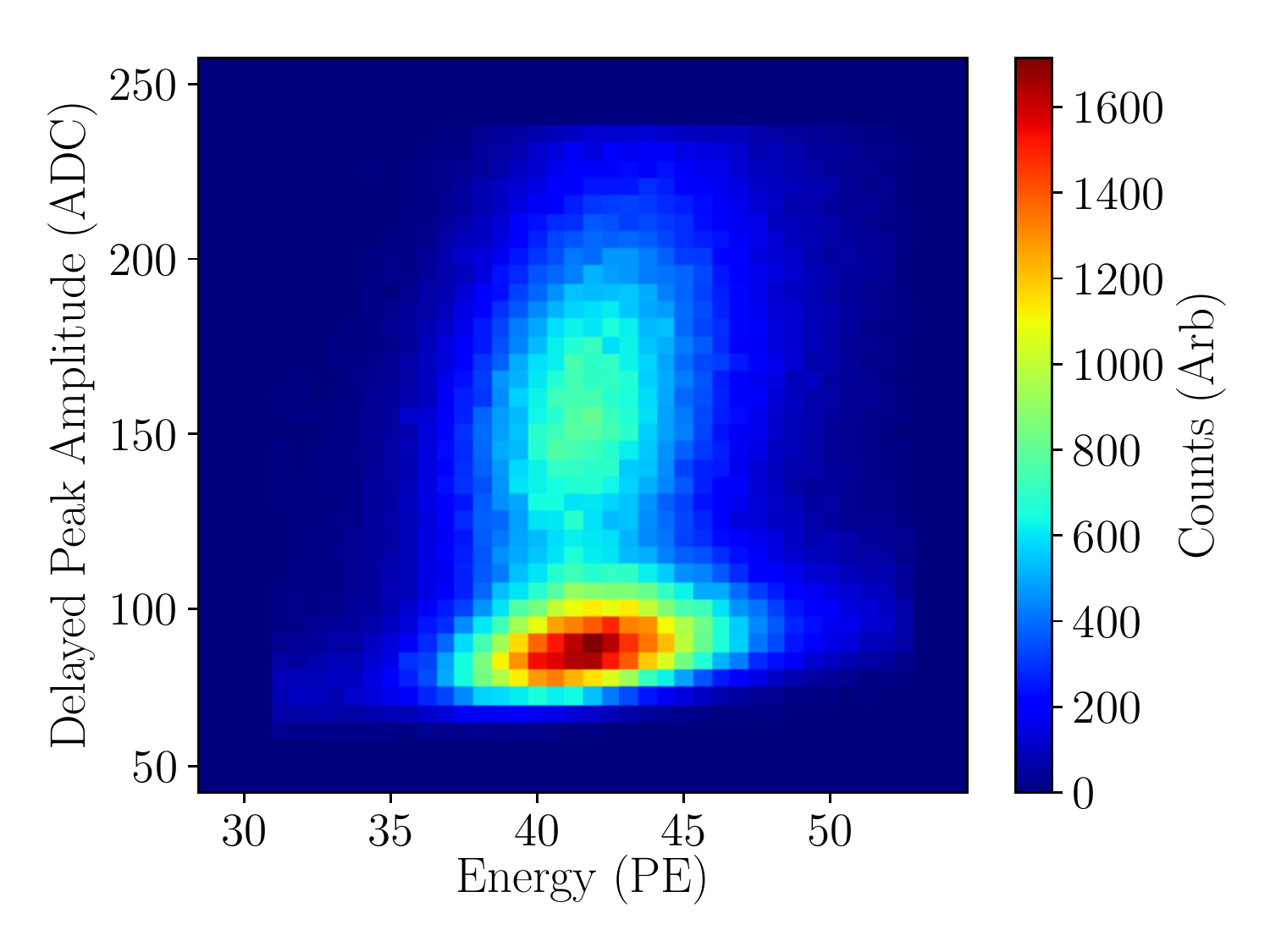}
\includegraphics[width=.48\textwidth]{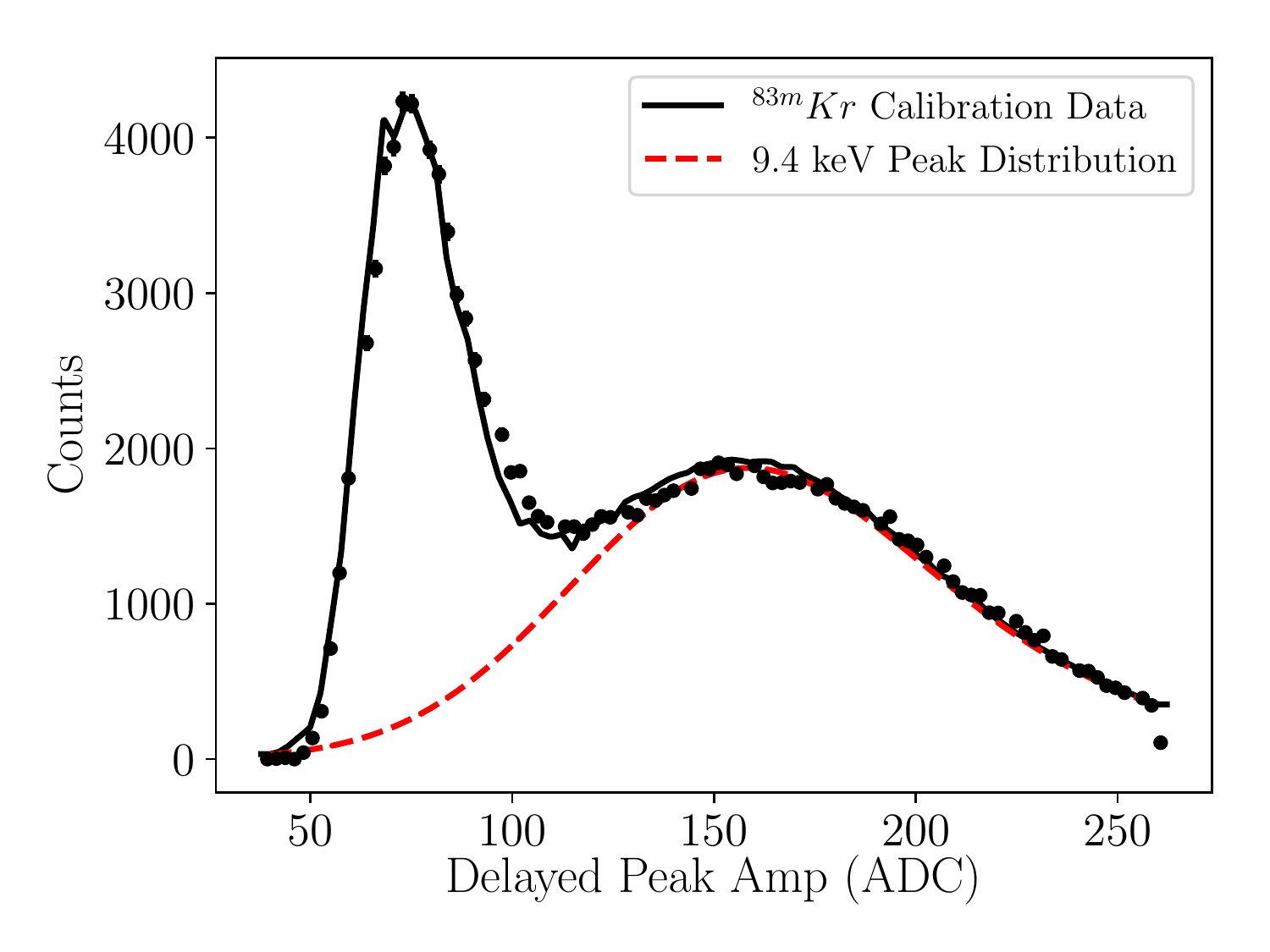}
    \caption{\textbf{Left}: Distribution of delayed peak amplitudes as determined through the peak-finding method described in the text. The population of delayed \SI{9.4}{\keV} pulses stands out clearly above the background. \textbf{Right}: Fit results for delayed peak amplitude \kr data. The solid black line represents the combined background and \SI{9.4}{\keV} distribution.}
    \label{fig:DelayedPeakAppearance}
\end{figure}

The peak height for \SI{9.4}{\keV} depositions was obtained through a peak-finding analysis of the summed PMT waveforms. After a delay of 152~ns with respect to event onset, the sample at which the largest amplitude occurs is selected as the ``delayed peak'' for that event. The peak information was limited to a single sample in this analysis to reduce the potential influence of triplet scintillation photons from the preceding \SI{32.1}{\keV} deposition. Examination of the timing distribution of these delayed peaks indicates that they originate from the \SI{9.4}{\keV} component as the number of detected secondary peaks decays in time with a half-life in agreement with the expected value of 157~ns for the \SI{9.4}{\keV} line in \kr, shown in Figure~\ref{fig:halflifetwo}.  Figure~\ref{fig:DelayedPeakAppearance} shows the clear appearance of the \SI{9.4}{\keV} deposition for events from a \kr calibration run.

To obtain a proper fit of the most probable peak height, the same peak-finding process was applied to background \arbeta data to build the underlying background distribution. Finally, the distribution of \SI{9.4}{\keV} peak heights was modelled by a normal distribution featuring an exponential tail. This model was combined with the \arbeta background distribution and a fit was performed on the \kr calibration data (shown in Figure~\ref{fig:DelayedPeakAppearance}), yielding a most probable peak height value of $158.5\pm1.4$~ADC\,units. This value was used in conjunction with Eq.~\eqref{eq:lyspline} to infer the light yield at \SI{9.4}{\keV}. Results from this method and the full integration method are shown in Table~\ref{tab:lycalibs}.

\begin{table}[!htbp]
\centering
\caption{\label{tab:lycalibs} Results of light yield calibrations for \kr components using the peak extrapolation methods.}
\smallskip
\begin{tabular}{l c c c}
\toprule
Energy (keV) & Method & $I$ (PE) & $Y$ (PE/keV)\\
\midrule
9.4 & Peak extrapolation  & 41.7 $\pm$ 2.2 & 4.44 $\pm$ 0.23 \\
32.1 & Peak extrapolation  & 135.7 $\pm$ 4.2 & 4.22 $\pm$ 0.13 \\
41.5 & Full integration & 176.8 $\pm$ 3.6 & 4.26 $\pm$ 0.09 \\
\bottomrule
\end{tabular}
\end{table}
\FloatBarrier

\subsubsection{Analysis B}

We also performed an alternative and complementary analysis that computed the light yield from the two \kr decay components using the methods from Analysis B further described in Ref.~\cite{Akimov:2020}. This analysis used the information contained in the first 90~ns of an event, mainly the singlet light from the argon scintillation for both the 32.1~keV and 9.4~keV components, to infer the full detector-photon yield from these components. The two types of events considered were those with only the singlet light from the 32.1~keV present in the first 90~ns and those that contain the singlet light from the full 41.5~keV. The position in time in the waveform of the second peak as shown in Figure~\ref{fig:krevent} distinguishes these classes of events. This is in contrast to the analysis described in Section~\ref{sec:compA}, which searched for the singlet light from the 9.4~keV events outside of the 90~ns singlet decay using the position of this pulse in the waveform. 

\begin{figure}[!htbp]
\center{\includegraphics[width=0.8\textwidth]{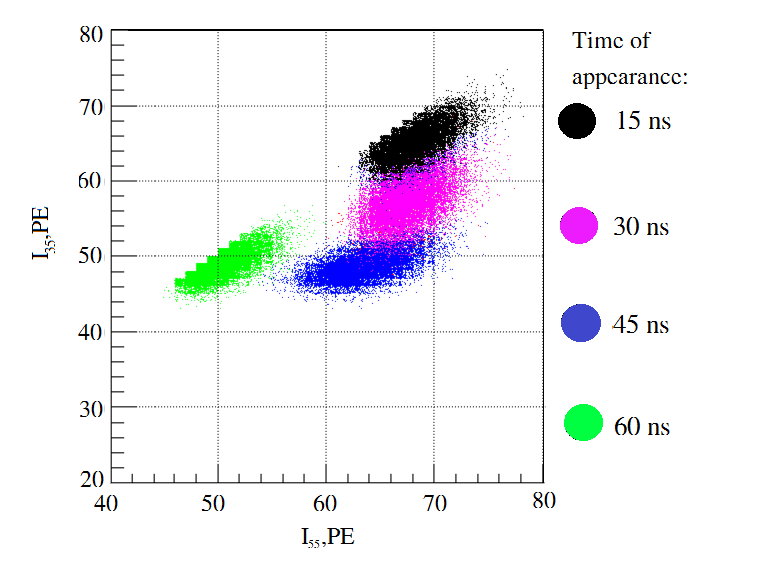}}
\caption{Distribution of the values of $I_{35}$:$I_{55}$ from the toy MC showing how the timing of the 9.4~keV component affects the distribution of these parameters. The different colors represent the various generated time separations between the 32.1~keV component and the 9.4~keV component.}
\label{fig:F35vsF55}
\end{figure}

Within this analysis, a time range of 0 to 55~ns selected events in which the singlet light from the 9.4~keV component was contained largely within the 90~ns with small losses of $<1\%$ in pulse integral. A toy Monte Carlo (MC) was performed to generate events in which the singlet light from the 9.4~keV component appeared a set time after the 32.1~keV component. Using the generated events from the MC, two parameters $I_{35}$ and $I_{55}$ were computed where $I_{35(55)}$ is the integral in the first 35(55)~ns of the waveform. The MC generated events with various time differences between the two components to study the behavior of $I_{35}$:$I_{55}$ in different populations. An example of this parameter space is shown in Figure~\ref{fig:F35vsF55}. The toy MC allows for an understanding of where these events will lie along a line by examining the two parameters based on the time at which the 9.4~keV component appears. If the singlet light from the 9.4~keV component occurs within the time range between 35 and 55 ns, then $I_{35} \neq I_{55}$. Events in the \cenns data can then be chosen in which the value lies below the $I_{35}\simeq I_{55}$ line in the $I_{35}$:$I_{55}$ parameter space in Figure~\ref{fig:F35vsF55}.  

\begin{figure}[!htbp]
\includegraphics[width=.48\textwidth]{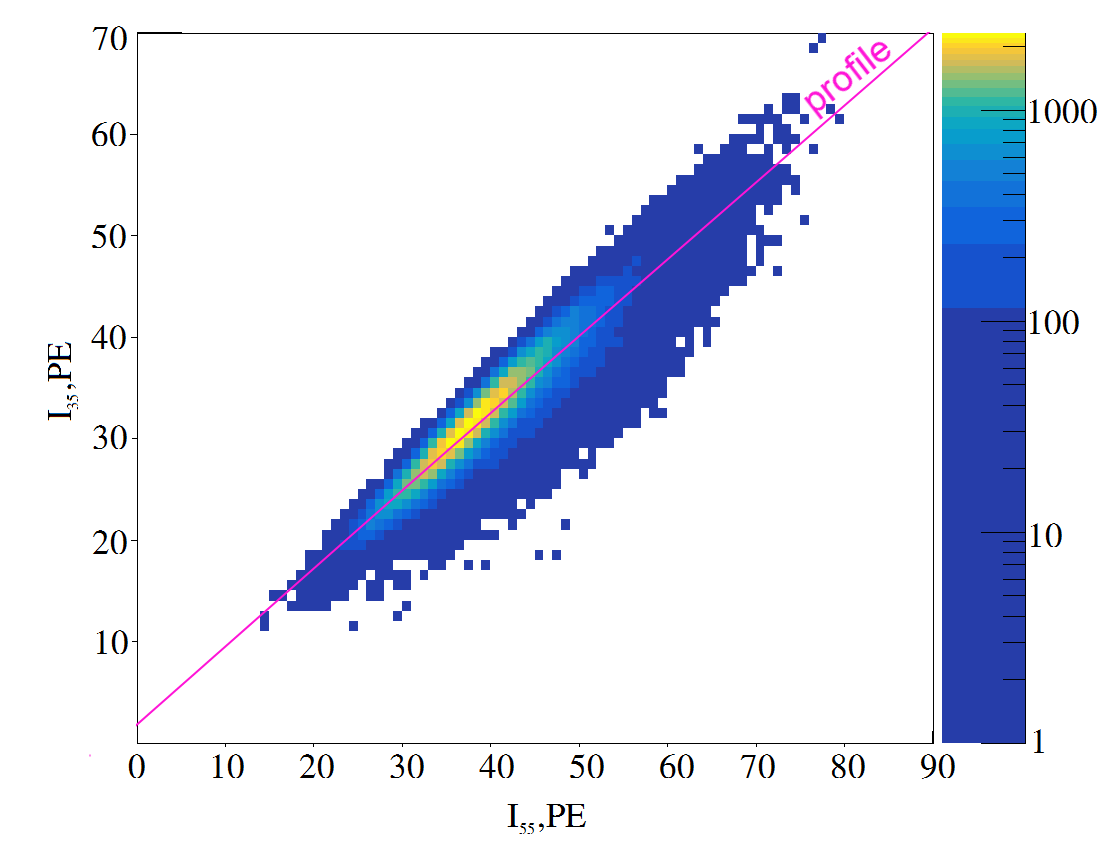}
\includegraphics[width=.48\textwidth]{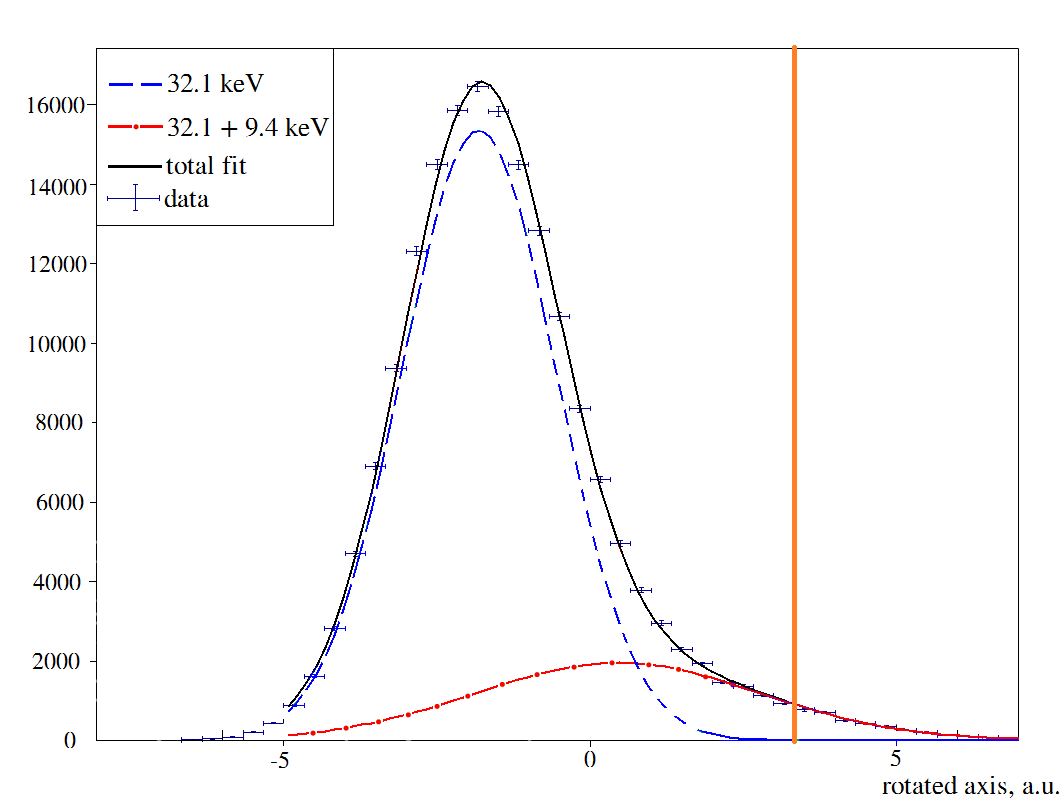}
    \caption{\textbf{Left}: Distribution of $I_{35}$:$I_{55}$ values in the \cenns \kr calibration data. Note the population below the line of $I_{35}\simeq I_{55}$ in the data showing events in which the 9.4~keV component is present inside the 35--55~ns window. \textbf{Right}: Projection of the data onto an axis that represents a rotated axis in the $I_{35}$ and $I_{55}$ space. The two Gaussians represent the two components and a cut was made to greatly suppress events in which the 9.4~keV component is not present to better select events with both components.}
    \label{fig:F35vsF55data}
\end{figure}

The same distribution of $I_{35}$ and $I_{55}$ values was found in the \cenns \kr calibration data. The left panel of Figure~\ref{fig:F35vsF55data} shows the calibration data with a population where $I_{35}\simeq I_{55}$ with some events lying below that line. The events below this line represent those in which the 9.4~keV component occurred between 35 and 55~ns after the 32.1~keV component and are the events of interest in this analysis. The data were projected onto an axis that represented a rotation of the $I_{35}$ and $I_{55}$ axes, which is shown in the right panel of Figure~\ref{fig:F35vsF55data}. The coordinate value on the rotated axis is given by
\begin{equation}
    \mathrm{rotated \: axis} = I_{35}\mathrm{cos}(\alpha) - I_{55}\mathrm{sin}(\alpha).
\end{equation}
A linear fit to the profile of the left panel of Figure~\ref{fig:F35vsF55data} along the $I_{55}$ axis gives $\alpha=37.5^{\circ}$ from the slope of the line. The two Gaussians represent events containing only the 32.1~keV component and both components, respectively. The vertical line represents a cut in the rotated axis, which suppresses the number of events with only the 32.1~keV component by a factor of $10^{3}$. 

Additionally, a cut was made on the value of $F_{90}=\frac{I_{90}}{I}$, the fraction of the total integral of the waveform in the first 90~ns. An $F_{90}$ cut selected \kr electron recoil events. $I_{90}$ represents the integrated signal in PEs in the first 90 ns of the waveform and $I$ represents the total integrated signal in PEs over the entire time window. The optimal cut was derived from the \co data, which also represent electron recoils but do not have a large probability of multiple recoils occurring in a single event window. A cut on $f_{top}$ was made to reduce edge effects when computing the light yield. 

The selected data were fit with two Gaussians making selecting events to the left of the cut shown in Figure~\ref{fig:F35vsF55data} for the 32.1~keV events and to the right of the cut shown in Figure~\ref{fig:F35vsF55data} for the events with both components present. The best-fit value of each Gaussian represents $I_{90}$ for events with only the 32.1~keV component present in the first 90~ns, and events with both components present in the first 90~ns. Subtracting the value of $I_{90}$ for the 32.1~keV energy deposition from the full 41.5~keV extracted the $I_{90}$ value for the 9.4~keV, shown in Table~\ref{tab:altsepanl}.   

\begin{table}[!htbp]
\centering
\caption{\label{tab:altsepanl} Results of light yield calibrations for \kr components using the peak extrapolation methods.}
\smallskip
\begin{tabular}{l c c}
\toprule
Energy ({keV}) & Method & $I_{90}$ (PE)\\
9.4 & Subtraction & 17.4 $\pm$ 0.2 \\
32.1  & Direct fit & 45.2 $\pm$ 0.6 \\
41.5 & Direct fit & 62.6 $\pm$ 1.1 \\
\bottomrule
\end{tabular}
\end{table}

We considered an iterative calculation approach applying the Levenberg--Marquardt\cite{Levenberg:1944, Marquardt:1963} algorithm to move from only considering the first 90~ns of the waveform to the full waveform. The algorithm solves the following system of equations to extract the total light yield for the two \kr components:
\begin{equation}
 \begin{split}
   I_{\mathrm{total},32.1} & = \frac{I_{90,32.1}}{(F_{90})(I_{32.1})}
   \\
%  \frac{FastLY_{32.1}}{f90(TotalLY_{32.1})} = TotalLY_{32.1} 
   I_{\mathrm{total},9.4} & = \frac{I_{90,9.4}}{(F_{90})(I_{9.4})}
   %\frac{FastLY_{9.4}}{f90(TotalLY_{9.4})} = TotalLY_{9.4}
   \\
   I_{\mathrm{total},9.4} & = I_{\mathrm{total},41.5} - I_{32.1}
   %TotalLY_{32.1+9.4} - TotalLY_{32.1} = TotalLY_{9.4}
   \\
   I_{\mathrm{total},32.1} & = I_{\mathrm{total}, 41.5} - I_{9.4}
   %TotalLY_{32.1+9.4} - TotalLY_{9.4} = TotalLY_{32.1}
   \\
   I_{90,9.4} & = I_{90,41.5} - I_{90,32.1}
   %FastLY_{32.1+9.4} - FastLY_{32.1} = FastLY_{9.4}
 \end{split}
\end{equation}
where, as above, $I_{90}$ represents the integral of the waveform in the first 90~ns, $I$ is the total integral of the waveform, and the $F_{90}$ parameters represent the values of $F_{90}$ extracted from the \co data at the values of $I_{\mathrm{total},32.1}$ and $I_{\mathrm{total},9.4}$. The subscripts 32.1 and 9.4 represent the two \kr components. The algorithm proceeds as follows:
\begin{itemize}
  \item The initial measured parameters given to the algorithm were $I_{90,32.1}$ and $I_{90,41.5}$ from the results in Table~\ref{tab:altsepanl}. The value of $I_{90,9.4}$ was also included but was allowed to float as a free parameter. $I_{\mathrm{total}, 41.5}$ was measured from the full \kr energy deposition. 
  \item Initial assumptions applying linearity to the 32.1~keV and 9.4~keV components computed $I_{\mathrm{total},32.1}$ and $I_{\mathrm{total},9.4}$. The calculation used the value of $Y$ shown in the right panel of Figure~\ref{fig:krspectrum} for Analysis B as the input to the system of equations.  
  \item The calculation iterated solving for the values of $I_{32.1}$, $I_{9.4}$, and $I_{90, 9.4}$ until convergence.
\end{itemize}
The results of the iterative calculation are given in Table~\ref{tab:itcalc}.
\begin{table}[!htbp]
\centering
\caption{\label{tab:itcalc} Results of iterative calculation computing the value of $Y$ for the two \kr components.}
\smallskip
\begin{tabular}{l c c}
\toprule
Energy (keV) & $I$ (PE) & $Y$ (PE/keV)\\
\midrule
9.4 & $44.7\pm6.3$ & $4.76\pm0.67$\\
32.1 & $149.2\pm9.8$ & $4.65\pm0.31$\\
\bottomrule
\end{tabular}
\end{table}

\subsubsection{Summary}
\begin{figure}[!htbp]
\centering
\includegraphics[width=.48\textwidth]{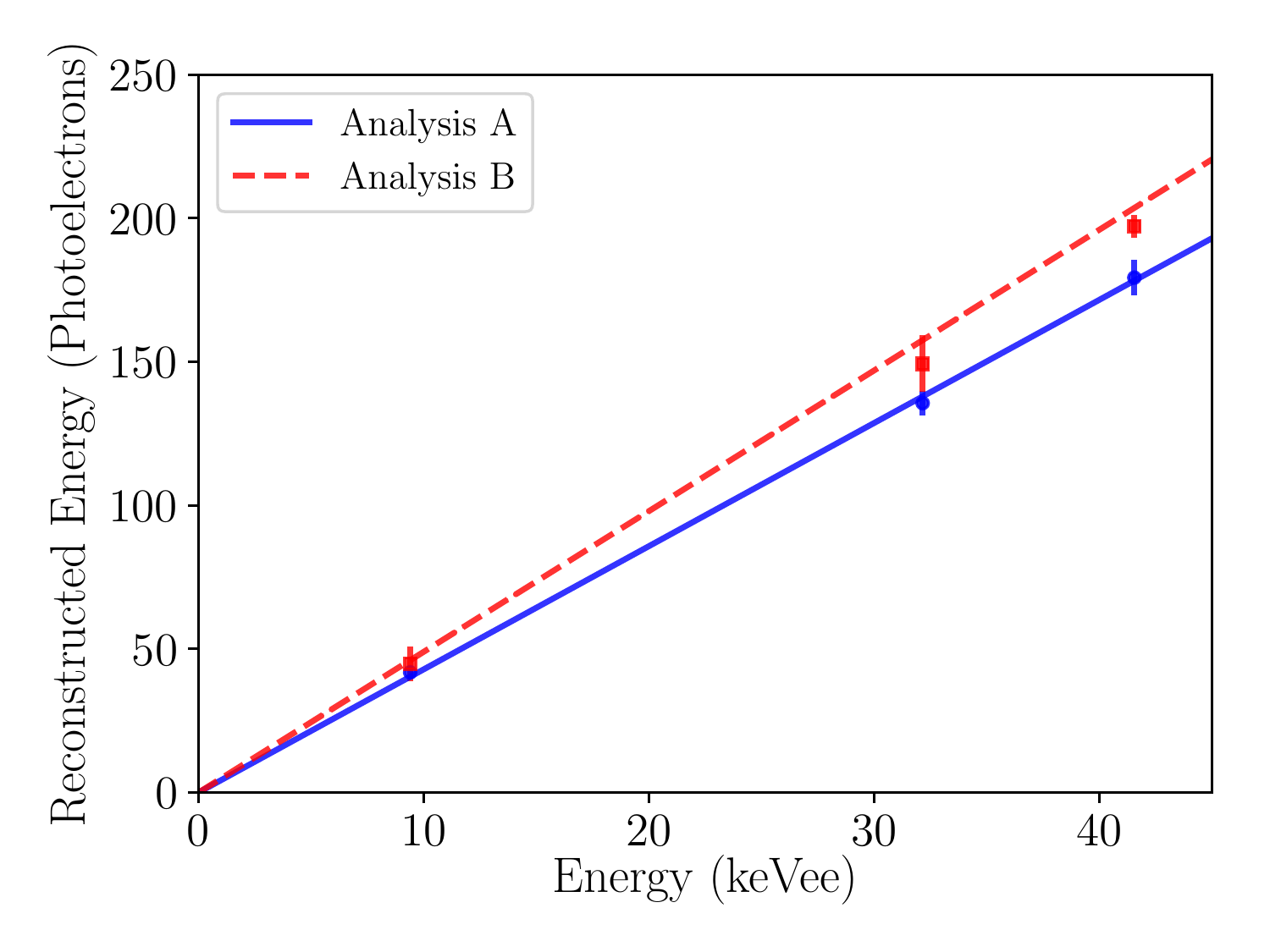}
\includegraphics[width=.48\textwidth]{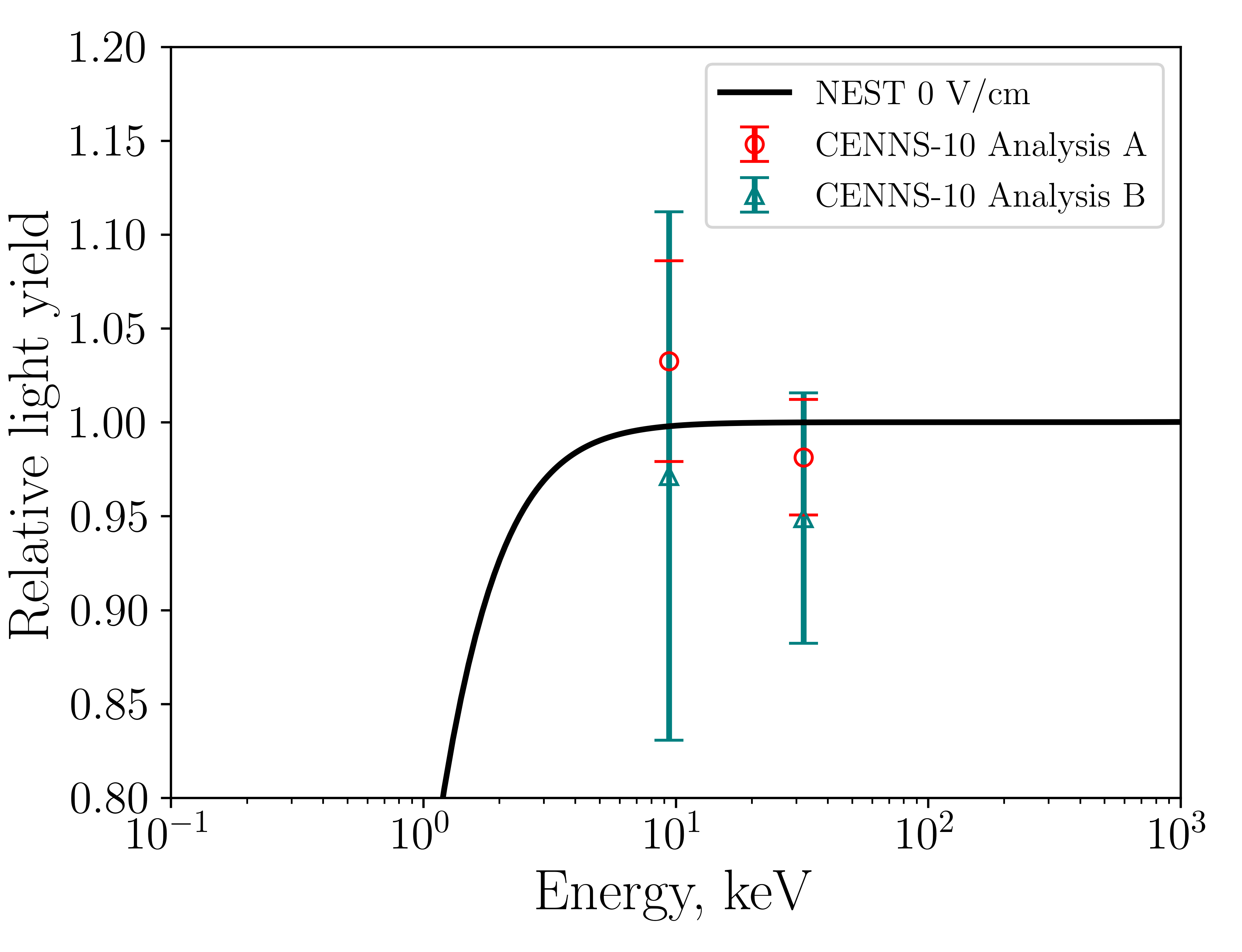}
\caption{\textbf{Left}: Measured detector response to the \kr source including the results from the component isolation analyses at 32.1~keV and 9.4~keV. The two fit lines correspond to the previously measured values of $Y=4.3\pm0.1$~PE/keV from Analysis A in blue and $Y=4.9\pm0.1$~PE/keV for Analysis B in red, shown in Figure~\ref{fig:krspectrum}. The results of these analyses agree with the linear detector response shown previously and push the measured linearity range to 9.4--122~keV. \textbf{Right}: Comparison of \cenns results from the two \kr component separation analyses compared with the ArgonNEST predictions. The results of these analyses agree with the prediction of detector response linearity from ArgonNEST.}
\label{fig:calib_total}
\end{figure}

The two analyses described in this section agree that the \cenns detector response was linear in the range 9.4 to 122~keV. The two methods were complementary, applying different techniques to extract the light yield for the two \kr components, and came to the same conclusion that detector response is linear. The results from Table~\ref{tab:lycalibs} for Analysis A and Table~\ref{tab:itcalc} for Analysis B are in agreement with a linear detector response when combined with the previous calibration results from Figure~\ref{fig:krspectrum}. This is shown graphically in the left panel of Figure~\ref{fig:calib_total}.  

Additionally, these results agree with the recent ArgonNEST~\cite{ArgonNEST:2020} release predicting linearity of liquid argon detectors at zero applied electric field down to 9.4~keV. A comparison of the results described in this section with the ArgonNEST predictions is shown in the right panel of Figure~\ref{fig:calib_total}. Because ArgonNEST predictions use the absolute photon yield from liquid argon, the relative light yield was computed with the \cenns results scaled by the light yield values for each analysis from the left panel of Figure~\ref{fig:calib_total}.

\section{Conclusions}
A \kr calibration source was designed and prepared for the calibration of COHERENT's \cenns liquid argon scintillation detector. The source was successfully integrated into the \cenns gas-handling and recirculation system. A measurement of the source activity was made using nondestructive assay techniques. The \kr evenly distributed within the detector $\sim30$~min after the initial introduction of \kr. A comparison of the measured activity to the steady-state \kr decay rate seen in \cenns showed that a sufficient fraction of the expected \kr decays reached the active volume, providing a calibration with good agreement between the measured \kr decay lifetime and the expected decay lifetime. \cenns operation in injection mode preserved stable detector operations and was the optimal method for calibration with \kr. 

The calibration runs taken with this source give confidence in continued calibration of the \cenns detector with \kr. A significant population of \kr decays appeared during a single calibration run. Analysis of \kr data shows the energy response of the detector was linear to \SI{9.4}{\keV}. The measured energy resolution of $\sim9\%$ was sufficient for a \cevns measurement with \cenns. The measured half-life of the \kr decay from the calibration runs agrees well with the expected value from the literature. Two additional independent analyses, which examine the two transitions in the decay of \kr, separately calculated the light yield from each transition. The results of the analyses agree with the recent ArgonNEST predictions of detector response linearity in this energy range. This source gave the lowest-energy calibration available to \cenns and was close to the energy region of interest for the primary \cevns search analysis. The results of the \kr calibration bolsters our understanding of the detector at the expected energies for \cevns recoil measurements.

\FloatBarrier

\acknowledgments
The authors would like to thank Richard Saldanha for useful discussions on the design and preparation of the \kr source. The COHERENT collaboration acknowledges the generous resources provided by the Oak Ridge National Laboratory Spallation Neutron Source, a DOE Office of Science User Facility, and thanks Fermilab for the continuing loan of the \cenns detector. This material is based upon work supported by the US Department of Energy (DOE), Office of Science, Office of Workforce Development for Teachers and Scientists, Office of Science Graduate Student Research (SCGSR) program. The SCGSR program is administered by the Oak Ridge Institute for Science and Education (ORISE) for the DOE. ORISE is managed by Oak Ridge Associated Universities under contract number DE-SC0014664. This work was supported by the Ministry of Science and Higher Education of the Russian Federation, Project ``Fundamental properties of elementary particles and cosmology'' No. 0723-2020-0041. We also acknowledge support from the Alfred P. Sloan Foundation, the Consortium for Nonproliferation Enabling Capabilities, the Institute for Basic Science (Korea, grant no. IBS-R017-G1-2019-a00), the National Science Foundation, and the Russian Foundation for Basic Research (projs. 20-02-00670\_a and 18-32-00910 mol a).  Laboratory Directed Research and Development funds from Oak Ridge National Laboratory also supported this project. This research used the Oak Ridge Leadership Computing Facility, which is a DOE Office of Science User Facility. This manuscript has been authored by UT-Battelle, LLC, under contract DE-AC05-00OR22725 with the US Department of Energy (DOE). The US government retains and the publisher, by accepting the article for publication, acknowledges that the US government retains a nonexclusive, paid-up, irrevocable, worldwide license to publish or reproduce the published form of this manuscript, or allow others to do so, for US government purposes. DOE will provide public access to these results of federally sponsored research in accordance with the DOE Public Access Plan \url{(http://energy.gov/downloads/doe-public-access-plan)}.

\FloatBarrier

\bibliographystyle{vitae}
\bibliography{bib}
\end{document}